\documentclass[11pt,a4paper]{article}
\usepackage[latin1]{inputenc}
\usepackage{amsmath}
\usepackage{amsfonts}
\usepackage{dsfont}
\usepackage{amssymb}
\usepackage{psfrag}
\usepackage{epsfig}
\usepackage{float}
\usepackage{hyperref}
\usepackage{caption}
\usepackage[dvipsnames]{xcolor}

\usepackage{tikz}
\usetikzlibrary{shapes.geometric, arrows}
\definecolor{hawaiianblue}{rgb}{0.4,0.68,0.953}
\definecolor{sand}{rgb}{0.863,0.953,0.811}
\definecolor{tropicalgreen}{rgb}{0.251, 0.857,0.341}

\tikzstyle{blob} = [rectangle, rounded corners, minimum width=3cm, minimum height=1cm,text centered, draw=black, fill=hawaiianblue!30, align = center]

\tikzstyle{blob2} = [rectangle, rounded corners, minimum width=3cm, minimum height=1cm,text centered, draw=black, fill=tropicalgreen!47, align = center]

\tikzstyle{blob3} = [rectangle, rounded corners, minimum width=3cm, minimum height=1cm,text centered, draw=black, fill=sand!51, align = center]

\tikzstyle{blob4} = [rectangle, rounded corners, minimum width=3cm, minimum height=1cm,text centered, draw=black, fill=Dandelion, align = center]

\tikzstyle{blob5} = [rectangle, rounded corners, minimum width=3cm, minimum height=1cm,text centered, draw=black, fill=hawaiianblue!30, align = center]

\tikzstyle{arrow} = [thick,->,>=stealth, align = center]

\newcommand{\nn}{\nonumber}
\newcommand{\ph}{\phantom}

\newcommand{\db}{D^{(\bar{\omega})}}
\newcommand{\bp}{\bar{\phi}}
\newcommand{\da}{\delta\alpha}

\title{A first-order approach to conformal gravity}

\author{T.G. Z\l o\'{s}nik$^{1,2}$\footnote{\texttt{zlosnik@fzu.cz}}\; and H.F. Westman$^3$\footnote{\texttt{hans.westman@alefomega.com}}
\\{\small \it $(1)$ Imperial College Theoretical Physics, Huxley Building, London, SW7 2AZ}
\\{\small \it $(2)$ 
	Institute of Physics of the Czech Academy of Sciences, Na Slovance 1999/2, 182 21, Prague}
\\{\small \it $(3)$ Alef Omega, Inc., 1023 Walnut St., 80302 Boulder, Colorado, United States}}
\begin{document}
\maketitle

\abstract{We investigate whether a spontaneously-broken gauge theory of the group $SU(2,2)$ may be a viable alternative to General Relativity. The basic ingredients of the theory are an $SU(2,2)$ gauge field $A_{\mu}$ and a Higgs field $W$ in the adjoint representation of the group with the Higgs field producing the symmetry breaking $SU(2,2)\rightarrow SO(1,3)\times SO(1,1)$. The action for gravity is polynomial in $\{A_{\mu},W\}$ and the field equations are first-order in derivatives of these fields. The new $SO(1,1)$ symmetry in the gravitational sector is interpreted in terms of an emergent local scale symmetry and the existence of `conformalized' General Relativity and fourth-order Weyl conformal gravity as limits of the theory is demonstrated. Maximally symmetric spacetime solutions to the full theory are found and stability of the theory around these solutions is investigated; it is shown that regions of the theory's parameter space describe perturbations identical to that of General Relativity coupled to a massive scalar field and a massless one-form field. 
The coupling of gravity to matter is considered and it is shown that Lagrangians for all fields are naturally gauge-invariant, polynomial in fields and yield first-order field equations; no auxiliary fields are introduced. Familiar Yang-Mills and Klein-Gordon type Lagrangians are recovered on-shell in the General-Relativistic limit of the theory. In this formalism, the General-Relativistic limit coincides with a spontaneous breaking of scale invariance and it is shown that this generates mass terms for Higgs and spinor fields.}

\section{Introduction}
The prevailing classical theory of gravity remains Einstein's General Relativity. The theory has enjoyed considerable success in accounting for data in gravitational experiments on scales of the solar system and below. Its success on larger scales is less clear. A considerable amount of evidence points towards the existence of a discrepancy between the properties of the universe as predicted by General Relativity in conjunction with known matter and what is actually observed \cite{Ade:2015xua}. The effects of this discrepancy can be incorporated into the framework of General Relativity via the introduction of additional matter described as a near-pressureless, perfect fluid (dark matter) but it is unclear whether this represents the effect of a genuinely new matter field or a manifestation of the breakdown of General Relativity. Additionally, there is evidence that the evolution of the very early universe was dominated by a scalar degree of freedom (the inflaton) that might not be formed from known matter and gravitational fields. Therefore, there exists an experimental motivation to consider alternative theories of gravitation \cite{Clifton:2011jh}.

In General Relativity the gravitational field is described solely by a field $e^{I} \equiv e^{I}_{\mu}dx^{\mu}$, the \emph{co-tetrad} \footnote{More usually gravity is discussed in terms of a metric tensor $g_{\mu\nu}\equiv \eta_{IJ}e^{I}_{\mu}e^{J}_{\nu}$ where $\eta_{IJ}= \mathrm{diag}(-1,1,1,1)$ is the invariant matrix of $SO(1,3)$. We choose to phrase things in terms of $e^{I}$ as it is only with this field that one can couple gravity to fermionic fields.}. The index $I$ denotes that the one-form $e^{I}$ is in the fundamental representation of the Lorentz group $SO(1,3)$. The action for General Relativity - the Einstein-Hilbert action - possesses an invariance under local Lorentz transformations represented by matrices $\Lambda^{I}_{\ph{I}J}(x)$ with $e^{I} \rightarrow \Lambda^{I}_{\ph{I}J}e^{J}$ \cite{Peldan:1993hi}. This field is sufficient to describe the inherent dynamics of gravity and the coupling of gravity to known matter fields:  scalar fields, gauge fields, and fermionic fields \footnote{Fermionic fields are spinorial representations of the group $SL(2,C)$ which is the double-cover of $SO(1,3)$ and so the inclusion of fermions into gravitational theory implies that most accurately the local symmetry of General Relativity is that of $SL(2,C)$.}.
In General Relativity the coupling of gravity to each of these fields requires use of the \emph{tetrad} $(e^{-1})_{I}\equiv e^{\mu }_{I}\partial_{\mu}$, where $e^{\mu}_{I}$ is the matrix inverse of $e_{\mu}^{I}$.  The coupling of gravity to all matter in General Relativity is hence non-polynomial and this contrasts with how other force fields in nature (the gauge fields of particle physics) couple to matter i.e. polynomial coupling within gauge-covariant derivatives of matter fields.
Such is the success of the standard model of particle physics, it is conceivable that an alternative to General Relativity would share its Lagrangian structure of fermionic, gauge, and Higgs fields coupled to one another polynomially; in Appendix \ref{gaugegravity} we review how a promising approach towards this arises from regarding the local $SO(1,3)$ symmetry of General Relativity to be the symmetry broken phase of a theory with a gauge symmetry of a larger group than the Lorentz group.

A more long-standing motivation to look at alternatives to General Relativity than the above considerations is the observation that General Relativity possesses fixed scales i.e. the Planck length and, potentially, the length scale provided by a cosmological constant; these scales may be interpreted as a kind of `absolute structure' in the theory. By comparison, the absolute time of Newton's theory of gravity was shown in the context of General Relativity to be an approximate notion arising from the - ultimately - dynamical spacetime geometry in the solar system. Does the presence of absolute scales in General Relativity indicate that it might be a limit of a larger theory in which such scales has a dynamical origin?  A model for the dynamical origin of scales was proposed by Zee \cite{Zee:1978wi} who introduced a scalar field $\Phi(x)$ such that the collection of scales in the gravitational sector should be due to this field having evolved to reach a fixed, non-zero value. A special case of such a theory is when 
a scalar field $\Phi(x)$ is coupled to gravity in a way such that field equations are invariant under the local field transformations $g_{\mu\nu} \rightarrow e^{2\alpha(x)}g_{\mu\nu}$ and $\Phi\rightarrow e^{-\alpha(x)}\Phi$; this is sometimes referred to as the conformal coupling of a scalar field to gravity. If the scalar field $\Phi$ in this case is non-vanishing then it can be set to a constant by a transformation with $\alpha = \ln\Phi$, in which case the gravitational part of the theory reduces to General Relativity \cite{Flanagan:2006ra}; thus $\Phi$ is `pure gauge' whenever it is non-vanishing. More generally, the invariance of a theory's field equations under these local transformations for the metric and other fields is referred to as \emph{local scale invariance} and a number of gravitational theories possessing this symmetry have been widely studied \cite{Boulware:1983td,Kazanas:1988qa,Sola:1988nz,Hehl:1989ij,Boulanger:2001he,Pireaux:2004xb,Bonezzi:2010jr}.

In this paper we will describe a theory of gravitation where gravity possesses a local $SU(2,2) \simeq SO(2,4)$ symmetry and with field content consisting of an $SU(2,2)$ gauge field $A_{\mu}$ and a Higgs field $W$ in the adjoint representation of the group which will be able to perform the symmetry breaking $SO(2,4)\rightarrow SO(1,3)\times SO(1,1)$. We will show that this structure allows gravity to couple polynomially to matter in a way more akin to the gauge fields of particle physics and that a consequence of pursuing this similarity is the existence of a new field in gravity: the gravitational Higgs field $W$; we take this field to be genuinely dynamical in the sense that it is not subject to any Lagrangian constraints.
This seems justified given the observation of the electroweak Higgs field which is in this sense unconstrained. It will be seen that the field $W$ may propagate, thus introducing new degrees of freedom into the gravitational sector, and the extent to which these new degrees of freedom may play a cosmological role is explored.
Furthermore, it will be shown that this theory contains scale-invariant models of gravity as limiting cases and may allow for a dynamical explanation for the origin of scale in the gravitational sector.

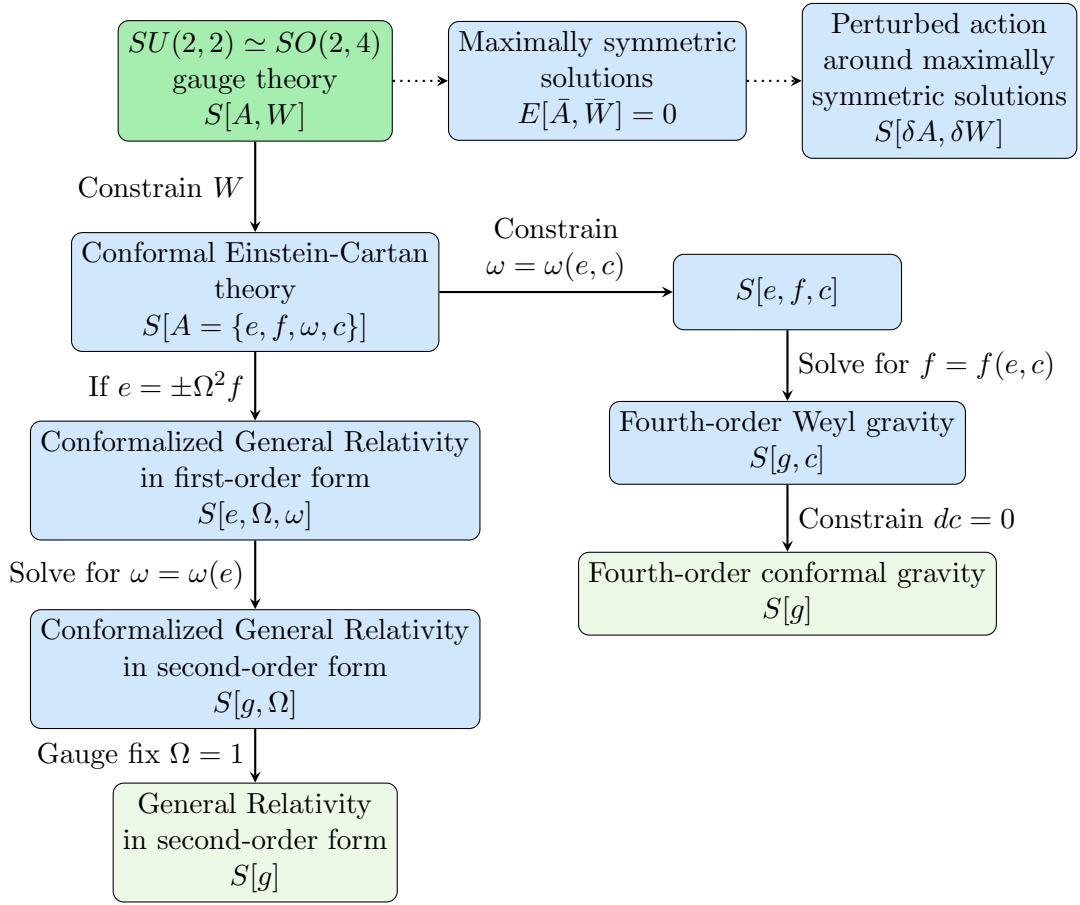
\begin{figure}[H]
	\begin{center}
		
		%% Nodes 
		\begin{tikzpicture}[node distance=2cm]
		\node (aw) [blob2] {$SU(2,2) \simeq SO(2,4)$ \\ gauge theory \\
			$S[A,W]$};
		\node(sol)[blob5, right of = aw, xshift = 2.5cm]{Maximally symmetric \\solutions \\ $E[\bar{A},\bar{W}]=0$};
		\node(solpert)[blob5, right of = sol,xshift = 2.5cm]{Perturbed action \\around  maximally\\ symmetric  solutions\\
			$S[\delta A,\delta W] $};
		\node (cec) [blob, below of=aw, yshift = -0.8cm] {Conformal Einstein-Cartan \\ theory \\ $S[A=\{e,f,\omega,c\}]$};
		\node (confgrav) [blob, left of=cec,xshift = 9cm] {$S[e,f,c]$};
		\node (confgr) [blob, below of=cec,yshift = -0.5cm] {Conformalized General Relativity \\ in
			first-order form\\
			$S[e,\Omega,\omega]$};
		\node (confgr2) [blob, below of=confgr,yshift = -0.5cm] {Conformalized General Relativity \\ in
			second-order form\\$S[g,\Omega]$};
		\node (confgr3) [blob3, below of=confgr2,yshift = -0.3cm] {General Relativity \\ in
			second-order form\\$S[g]$};
		\node(weylg)[blob, below of =confgrav]{Fourth-order Weyl gravity\\ $S[g,c]$};
		\node(weylg2)[blob3,below of = weylg]{Fourth-order conformal gravity \\ $S[g]$};
		%% Arrows
		
		\draw [dotted,arrow] (aw) -- (sol);
		\draw [dotted,arrow] (sol) -- (solpert);
		\draw [arrow] (aw) --node[anchor = east]{Constrain  $W$} (cec) ;
		\draw [arrow] (cec) --node[anchor= south]{Constrain \\  $\omega=\omega(e, c)$} (confgrav) ;
		\draw [arrow] (cec) --node[anchor =east]{If  $e = \pm \Omega^2 f$} (confgr);
		\draw [arrow](confgrav) --node[anchor = west]{Solve for $f=f(e,c)$ } (weylg);
		\draw [arrow](confgr) --node[anchor=east]{Solve for  $\omega=\omega(e)$} (confgr2);
		\draw[arrow](confgr2) --node[anchor =east]{Gauge fix $\Omega = 1$} (confgr3);
		\draw[arrow](weylg) -- node[anchor = west]{Constrain $dc=0$} (weylg2);
		
		\end{tikzpicture}
		
	\end{center}
	
	\captionsetup{width=0.89\textwidth, font = small}
	\caption{Diagram depicting known results of $SU(2,2) \simeq SO(2,4)$ gravity and relation to other gravitational models. Dotted paths denote those taken with $W^{AB}$ entirely unconstrained.}
	\label{fig2}
	
\end{figure}

The outline of the paper is as follows: In Section \ref{propact} we discuss the field content of the theory and present its action. In Section \ref{cec} we consider the limit of the theory that follows from  constraining all of the degrees of freedom present in the Higgs field $W$ and show how fourth-order conformal gravity \cite{MannheimKazanas1989} and a conformally coupled scalar tensor theory arise. It is shown that in this limiting theory scale symmetry is broken not by the introduction of a new scalar field but by the dynamical alignment of two sets of frame fields. In this section we also discuss motivations for looking at the group $SO(2,4)$ as an approach to conformal gravity rather than other orthogonal groups such as $SO(1,5)$ or $SO(3,3)$. In Section \ref{maxsym} we return to the full theory and show that there exist maximally symmetric solutions to the field equations i.e. solutions interpretable as spacetimes possessing ten Killing vectors each. In Section \ref{perts} we examine the nature of small perturbations around these solutions, establishing conditions for linear stability. To aid the reader, the flow chart Figure \ref{fig2} summarizes the structure of these sections and some results therein. In Section \ref{matter} we discuss how one can couple the gravitational fields to matter fields in a simple and elegant fashion. All actions are polynomial and yield equations of motion that are first order partial differential equations that in the General-Relativistic limit of the theory reduce to the familiar second order equations for Yang-Mills fields and Higgs fields. It is suggested that Yang-Mills fields for a symmetry group ${\cal G}$ are necessarily accompanied by scalar fields in the adjoint representation of ${\cal G}$.  In Section \ref{observable} we discuss some potentially observable consequences of the model with a focus on open problems in cosmology. In Section \ref{other} we discuss the relation of the work presented in this paper to previous approaches in the literature and in Section \ref{disc} we discuss the paper's results and present conclusions.

%% Flow diagram
 
 \section{Gravitational action}
 \label{propact}
The group $SU(2,2)$ has a matrix representation as the set of all $4\times 4$ complex matrices $U^{\alpha}_{\ph{\alpha}\beta}$ \footnote{We reserve $\alpha,\beta,\gamma,\delta$ as $SU(2,2)$ indices whereas other Greek letters will be used to denote spacetime indices.} of unit-determinant that satisfy

\begin{eqnarray}
h_{\alpha\alpha'} =  h_{\beta\beta'}U^{\beta}_{\ph{\beta}\alpha}U^{*\beta'}_{\ph{\beta'}\alpha'}, \quad  h_{\alpha\alpha'}  =   \left( \begin{array}{cc}
0 & I \\
I &0
\end{array} \right)
\end{eqnarray}
where $I$ is the $2\times2$ identity matrix. This group is the double cover of the orthogonal group $SO(2,4)$ which itself is the double cover of the \emph{conformal group} $C(1,3)$ of coordinate transformations which preserve a metric of signature $(-,+,+,+)$ up to an overall multiplicative function. In taking gravity to possess a local $SU(2,2)$ symmetry we imply that gravitational fields and fields coupling to gravity belong to representations of this group and that terms in the Lagrangian involving these fields are invariant under local $SU(2,2)$ transformations $U^{\alpha}_{\ph{\alpha}\beta}(x)$. The group $SU(2,2)$ is fifteen-dimensional whilst the Lorentz group $SO(1,3)$ associated with General Relativity is six-dimensional so it is necessary for spontaneously symmetry breaking to occur in order to recover a General-Relativistic limit of the theory. We will show that a suitable Higgs sector for the $SU(2,2)$ theory can break the $SU(2,2)$ symmetry down to $SO(1,3)\times SO(1,1)$; this is a larger symmetry than General Relativity and we will show how it is related to local scale-invariance.

For calculational purposes it is rather convenient in the gravitational sector to work with representations of $SO(2,4)$. The machinery to move from representations of $SU(2,2)$ to $SO(2,4)$ is discussed in detail in Section \ref{spinors} when we discuss the coupling of gravity to fermions, which we take to be fields in the fundamental representation of $SU(2,2)$. A group element of $SO(2,4)$ may be represented as a matrix $\Lambda^{A}_{\ph{A}B}$ with unit-determinant and satisfying

\begin{eqnarray}
\eta_{AB} =  \eta_{CD}\Lambda^{C}_{\ph{C}A}\Lambda^{D}_{\ph{D}B} , \quad \eta_{AB} = \mathrm{diag}(-1,-1,1,1,1,1)
\end{eqnarray}
A field in the fundamental representation of $SO(2,4)$ is a six-component vector $U^{A}$ and 
a field in the adjoint representation takes of the form of an antisymmetric matrix $Y^{AB} = -Y^{BA}$. Indices are lowered and raised with $\eta_{AB}$ and its matrix inverse. Under a local $SO(2,4)$ transformation, the field $U^{A}$ transforms as $U^{A} \rightarrow \Lambda^{A}_{\ph{A}B}(x)U^{B}$ and we require that our Lagrangians are invariant under such transformations for any fields belonging to representations of $SO(2,4)$.

 The gravitational fields will be an $SO(2,4)$ gauge field $A^{AB}= A_{\mu}^{\ph{\mu}AB}dx^{\mu}$ and a spacetime scalar field $W^{AB}$ in the adjoint representation. The field $W^{AB}$ can always be put in the following `block-diagonal' form by appropriate $SO(2,4)$ transformations:

 \begin{equation}
 W^{AB} = \left( \begin{array}{ccc}
t_{1}\Sigma & 0 & 0\\
0 & t_{2}\Sigma& 0\\
0&0 & t_{3}\Sigma \end{array} \right) , \quad \Sigma \equiv \left( \begin{array}{cc}
0 & 1 \\
-1 &0
 \end{array} \right)  \label{blockdiag}
 \end{equation}
If $t_{1}= t_{2}= 0$ (labeling indices on their rows and columns by $I,J,K,\dots = 0,1,2,3$) and $t_{3}\neq 0$ (labeling indices on its rows and columns by $a,b,c,\dots = -1,4$) then this form of $W^{AB}$ will be invariant under the $SO(2,4)$ transformations  $\Lambda^{I}_{\ph{I}J}$ and $\Lambda^{a}_{\ph{a}b}$. If $\mathrm{sign}(\eta_{-1-1}) \neq \mathrm{sign}(\eta_{44})$ then - by implication - $\Lambda^{a}_{\ph{a}b}$ represent hyperbolic rotations/boosts in the $(-1,4)$ plane whilst $\Lambda^{I}_{\ph{I}J}$ are transformations generated by the Lorentz group subgroup of $SO(2,4)$. Thus with $t_{3} \neq 0$, and $t_{1},t_{2}=0$,  the residual symmetry is $SO(1,3)\times SO(1,1)$. 

Useful quantities are the curvature two-form  $F^{AB}$ and covariant derivative one-form of $W^{AB}$ - $DW^{AB}$ - given as follows:

\begin{eqnarray}
F^{AB} &=&  dA^{AB} +  A^{AC}A_{C}^{\ph{C}B}\\
DW^{AB} &=& dW^{AB} + A^{AC}W_{C}^{\ph{C}B} + A^{BC}W^{A}_{\ph{A}C}
\end{eqnarray}
where $d$ is the exterior derivative on forms and multiplication of forms is always taken to be via the wedge product. Note that for differential forms $a,b,c$ the wedge product satisfies $a(b+c) = ab+ac$.
We will look to consider the most general locally $SO(2,4)$-invariant and diffeomorphism-invariant action polynomial in $\{A^{AB},W^{AB}\}$. We will take actions to be integrals of Lagrangian four-forms and our restriction to Lagrangians which are diffeomorphism-invariant and polynomial in fields is a simplifying principle. We may then build four-forms from $F^{AB}$ and $DW^{AB}$, thus guaranteeing that the Lagrangian is coordinate-independent. To further enforce local $SO(2,4)$ invariance, we will look to contract away all free $SO(2,4)$ indices for which we can in principle additionally use the scalar $W^{AB}$ and the $SO(2,4)$ invariants $\eta_{AB}$ and the completely antisymmetric symbol $\epsilon_{ABCDEF}$. Our action is as follows:

\begin{eqnarray}
S[A,W]&=&  \int  a_{ABCD}F^{AB}F^{CD}  + b_{ABCD}DW^{A}_{\ph{A}E}DW^{EB}F^{CD}
\nn \\
&&+ c_{ABCD} DW^{A}_{\ph{A}E}DW^{EB}DW^{C}_{\ph{C}F}DW^{FD} \label{act5}
\end{eqnarray}
where:
\begin{eqnarray}
a_{ABCD} &\equiv &  a_{1}\epsilon_{ABCDEF}W^{EF}+a_{2}W_{AE}W^{E}_{\ph{A}D}\eta_{BC}+a_{3}W_{AB}W_{CD}\nn\\
&& +  a_{4} \eta_{AC}\eta_{BD}+a_{5}\eta_{AC}W_{BD} \\
b_{ABCD} &\equiv & b_{1} \epsilon_{ABCDEF}W^{EF} \\
c_{ABCD} &\equiv &  c_{1} \epsilon_{ABCDEF}W^{EF}
\end{eqnarray}
The zero-form coefficients $\{a_{i},b_{1},c_{1}\}$ may in general depend on $SO(2,4)$ invariants built from $W^{AB}$ and the group invariants $\eta_{AB}$, $\epsilon_{ABCDEF}$. In this paper, as a first approach to the theory, we will take these coefficients to be constant numbers but it is conceivable that functional dependences on such invariants cannot be consistently neglected.
Though the action contains terms quadratic in $F^{AB}$ and quartic in $DW^{AB}$, the wedge product structure guarantees that \emph{components} of these fields appear at most linearly in the action. The generation of higher-order partial derivatives in the equations of motion is made impossible by use of a polynomial Lagrangian and  Bianchi identities $DF^{AB}=0$ and `$DDV=FV$'. Therefore, as the Lagrangian is at most linear in derivatives of any component, the equations of motion are at most first order in derivatives. The action may look very unfamiliar and so in the next section we will initially look at a simpler theory emerges when we `freeze' all the degrees of freedom in the field $W^{AB}$.  Finally, we note that though we will use $W^{AB}$ in the following sections,  we may alternatively (and equivalently) use an entirely antisymmetric field  $Y_{ABCD}$,  where the two are related via 

\begin{eqnarray}
Y_{ABCD} =\frac{1}{2} \epsilon_{ABCDEF}W^{EF}
\end{eqnarray}

\section{Conformal Einstein-Cartan theory}
\label{cec}

We first discuss a theory which emerges when the degrees of freedom of $W^{AB}$ in the model (\ref{act5}) are completely frozen by means of constraints imposed at the level of the action. Recall that the theory recovered by the same process of freezing the Higgs degree of freedom for the $SO(1,4)|SO(2,3)$ gauge theories resulted in the Einstein-Cartan theory, which in the absence of matter is equivalent to General Relativity (see Appendix \ref{gaugegravity}). We shall show that the corresponding theory for gravity based on the gauge group $SU(2,2)\simeq SO(2,4)\simeq C(1,3)$ is in some senses a straightforward scale-invariant generalization of the Einstein-Cartan theory.

Assuming the block-diagonal form of $W^{AB}$ from (\ref{blockdiag}), degrees of freedom in $W^{AB}$ can be characterized in terms of three $SO(2,4)$-invariant quantities:

\begin{eqnarray}
{\cal C}_{1} &=& \epsilon_{ABCDEF}W^{AB}W^{CD}W^{EF} = 6t_{1}t_{2}t_{3} \\
{\cal C}_{2} &=& W_{AB}W^{AB} =  -2\left(\eta_{00}\eta_{11}(t_{1})^{2}+\eta_{22}\eta_{33}(t_{2})^{2}+\eta_{-1-1}\eta_{44}(t_{3})^{2}\right) \\
{\cal C}_{3} &=& W_{AB}W^{B}_{\ph{B}C}W^{CD}W_{D}^{\ph{D}A} = 2\left((t_{1})^{4}+(t_{2})^{4}+(t_{3})^{4}\right)
\end{eqnarray}
If it is enforced that ${\cal C}_{1}=0$ and ${\cal C}_{3}=({\cal C}_{2})^{2}$, this implies that two out of $\{t_{1},t_{2},t_{3}\}$ are zero. For example, we can choose $t_{3}=0$ (the label `3' is completely arbitrary at this point) from ${\cal C}_{1}=0$. Subsequently, the condition $({\cal C}_{2})^{2}-{\cal C}_{3}=0$ takes the form:

\begin{eqnarray}
-8\eta_{00}\eta_{11}\eta_{22}\eta_{33}(t_{1})^{2}(t_{2})^{2}=0
\end{eqnarray}
Hence we can choose $t_{2}=0$. Now, if we further require ${\cal C}_{2}>0$ we have the condition:

\begin{eqnarray}
-2\eta_{00}\eta_{11}(t_{1})^{2} >0
\end{eqnarray}
Therefore $\eta_{00}\eta_{11}<0$ and so we have the breaking of the original $SO(2,4)$ symmetry down to $SO(1,3)\times SO(1,1)$. 
If we then add on the following four-form Lagrange-multiplier constraints to the theory:

\begin{eqnarray}
S_{\lambda}[W^{AB},\lambda_{1},\lambda_{2},\lambda_{3}] &=& \int \lambda_{1}{\cal C}_{1} + \lambda_{2}\left({\cal C}_{2} - \bp^{2}\right)+\lambda_{3}\left({\cal C}_{3}-({\cal C}_{2})^{2}\right)
\end{eqnarray}
then the constraints will be enforced via the field equations obtained by varying $\lambda_{i}$. We may instead enforce the constraints at the level of the action, and so $W^{AB}$ may be assumed to take the following form at the level of the action:

 \begin{equation}
W^{AB} \overset{*}{=} \left( \begin{array}{ccc}
0 & 0 & 0\\
0 & 0& 0\\
0&0 & \bp \epsilon^{ab} \end{array} \right)\label{mat1}
\end{equation}
where $\overset{*}{=}$ implies an equality that holds in a specified gauge, we use indices $a,b,c\dots$ as indices in the fundamental representation of  $SO(1,1)$, we use the convention $\epsilon_{-14} = 1$,  $\epsilon^{-14} = -1$ and the quantity $\bp$ is a constant. Given the application of these constraints, there are no longer any degrees of freedom for $W^{AB}$ left in the action (\ref{act5}); the action is now a functional only of $A^{AB}$, a general ansatz for which is given in this gauge by:

 \begin{equation}
 A^{AB} \overset{*}{=} \left( \begin{array}{cc}
\omega^{IJ} & E^{Ia} \\
-E^{Ia} & c \epsilon^{ab}
 \end{array} \right) \label{aab2}
 \end{equation}
The one-form field $\omega^{IJ}$ is the Lorentz-group spin connection, while the one-form $c$ is a connection for the group $SO(1,1)$. The `off-diagonal' components $E^{Ia}$ look much less familiar;  they transform homogeneously under the remnant $SO(1,3)\times SO(1,1)$ symmetry and appear in the $SO(2,4)$-covariant derivative of $W^{AB}$ as follows:

 \begin{equation}
 DW^{AB}  \overset{*}{=}  \left( \begin{array}{cc}
0 & \bp E^{I}_{\ph{I}c}\epsilon^{ca} \\
-\bp E^{I}_{\ph{I}c}\epsilon^{ca} & 0 \label{dwab}
 \end{array} \right)\end{equation}
Now we write down the total constrained form of (\ref{act5}) in the `preferred gauge', making use of the following results:

\begin{equation}
F^{AB}  \overset{*}{=}    \left( \begin{array}{cc}
R^{IJ}-  E^{I}_{\ph{I}a}E^{Ja} &  D^{(\omega+c)}E^{Ia} \\
-  D^{(\omega+c)}E^{Ia}&  dc \epsilon^{ab} - E_{J}^{\ph{J}a}E^{Jb}
 \end{array} \right)
, \quad DW^{A}_{\ph{A}C}DW^{CB}  \overset{*}{=}    \left( \begin{array}{cc}
\bp^{2} E^{I}_{\ph{I}d}E^{Jd} &  0 \\
0&  0
 \end{array} \right)
\end{equation}
where $R^{IJ} \equiv d\omega^{IJ}+\omega^{I}_{\ph{I}K}\omega^{KJ}$ is the $SO(1,3)$ curvature two-form; the action then becomes:

\begin{align}
S[\omega,c,E] &\overset{*}{=}  \int 2\bp( 2a_{1} -  \bp^{2}b_{1})\epsilon_{IJKL}E^{I}_{\ph{I}a}E^{Ja}R^{KL}
-2\bp(a_{1}-b_{1}\bp^{2}+c_{1}\bp^{4})\epsilon_{IJKL}E^{I}_{\ph{I}a}E^{Ja}E^{K}_{\ph{K}b}E^{Lb}
\nn \\
&  +a_{2}\bp^{2}R_{IJ}E^{Ja}E^{I}_{\ph{I}a}
+\bp^{2}(a_{2}+2a_{3})dc\epsilon^{cd}E_{Jc}E^{J}_{\ph{J}d}  + a_{2}\bp^{2} E_{Jd}E^{J}_{\ph{J}c}E_{K}^{\ph{K}c}E^{Kd} \nn\\
&+ a_{3}\bp^{2}\epsilon_{ab}\epsilon_{cd} E_{J}^{\ph{J}a}E^{Jb}E_{K}^{\ph{K}c}E^{Kd} +2\bp^{2}(a_{2}+2a_{3})dcdc  -2\bp a_{1}\epsilon_{IJKL}R^{IJ}R^{KL}  \label{cec1}
\end{align}
where we have used the result $DDE^{Ia} = R^{I}_{\ph{I}J}E^{Ja} + dc \epsilon^{a}_{\ph{a}b}E^{Ib}$. To make further progress, we make the following general ansatz for $E^{Ia}$ in an arbitrarily chosen $SO(1,1)$ gauge:

\begin{align}
E^{Ia} & \overset{*}{=}  \left(
\begin{array}{c}
e^{I}\\
f^{I}\\
\end{array}
\right) \label{eanz}
\end{align}
where components in the column vector (\ref{eanz}) are in components in an arbitrary null basis of the space of $SO(1,1)$ vectors. If we transform from this basis to another basis with an $SO(1,1)$ transformation represented by a matrix

 \begin{equation}
\Lambda^{a}_{\ph{a}b}  = \frac{1}{\sqrt{1-v^{2}}}\left( \begin{array}{cc}
1 & -v \\
-v & 1 \label{dwab}
\end{array} \right)\end{equation}
Then in the new basis we have

\begin{align}
E'^{Ia} & = \left(
\begin{array}{c}
e'^{I}\\
f'^{I}\\
\end{array}
\right) =  \left(
\begin{array}{c}
e^{-\beta}e^{I}\\
e^{\beta}f^{I}\\
\end{array}
\right) \\
c' &= c - d\beta
\end{align}
where $\beta \equiv \ln\left(\sqrt{\frac{1+v}{1-v}}\right)$. Thus, $SO(1,1)$ transformations cause dilations of opposing weight in $e^{I}$ and $f^{I}$. Applying the ansatz (\ref{eanz}) to (\ref{cec1}) yields:

\begin{eqnarray}
S[\omega,c,e,f] &\overset{*}{=}& \int \frac{1}{32\pi \tilde{G}}\left(\epsilon_{IJKL}e^{I}f^{J}R^{KL}-\epsilon_{IJKL}\frac{\tilde{\Lambda}}{6}e^{I}f^{J}e^{K}f^{L}- \frac{2}{\gamma}e^{I}f^{J}R_{IJ} \right)\nn\\
&& + \xi\left( \frac{1}{2}e_{I}f^{I}e_{J}f^{J}-dc e^{I}f_{I} \right) +{\cal C}_{1}\epsilon_{IJKL}R^{IJ}R^{KL}+ {\cal C}_{2} dcdc \label{cec4}
\end{eqnarray}
where
\begin{eqnarray*}
\frac{1}{16\pi \tilde{G}}(\bp) &=&  4\bp (2a_{1}-b_{1}\bp^{2}), \quad \tilde{\Lambda}(\bp) =  \frac{6(a_{1}-b_{1}\bp^{2}+c_{1}\bp^{4})}{(2a_{1}-\bp^{2}b_{1})} , \quad \gamma(\bp) =\frac{4(2a_{1}-b_{1}\bp^{2})}{a_{2}\bp}\nn \\
\xi(\bp) &=&  (a_{2}+2 a_{3})\bp^{2} , \quad {\cal C}_{1}(\bp) =  -2\bp a_{1} , \quad {\cal C}_{2}(\bp)= 2(a_{2}+2a_{3})\bp^{2}
\end{eqnarray*}
As expected, the action is manifestly locally Lorentz invariant and invariant under local $SO(1,1)$ gauge transformations $c\rightarrow c - d\beta(x)$, $e^{I}\rightarrow e^{-\beta(x)}e^{I}$, $f^{I}\rightarrow e^{\beta(x)}f^{I}$. The action additionally possesses invariance under local dilations of $e^{I}$ and $f^{I}$ of opposite weight:

 \begin{eqnarray}
 e^{I} \rightarrow e^{-\alpha(x)}e^{I},\quad  f^{I} \rightarrow e^{\alpha(x)} f^{I}
 \end{eqnarray} 
where $\alpha(x)$ is entirely independent gauge transformation parameter $\beta(x)$. By appearance, the theory (\ref{cec4}) resembles the Einstein-Cartan theory but instead has a \emph{pair} of frame fields $\{e^{I},f^{I}\}$ with a local scale symmetry under opposite rescalings due to them always appearing in the combination $e^{I}f^{J}$ in the Lagrangian. Note that this is only possible because of the choice of the group $SO(2,4)$ which led to a remnant $SO(1,1)$ internal symmetry in the gravitational sector which in turn allowed for the definition of two frame fields in terms of a \emph{null} basis in the space of $SO(1,1)$ vectors. If, instead, we had looked at theories with a local gauge symmetry for the de Sitter groups $SO(1,5)$ and $SO(3,3)$, an adjoint Higgs field could break the symmetry instead to $SO(1,3)\times SO(2)$; one could proceed to choose an arbitrary basis for the $SO(2)$ vector space, so defining frame fields $\{{\cal E}^{I},{\cal F}^{I}\}$ in the manner that $\{e^{I},f^{I}\}$ were defined in (\ref{eanz}). It can be shown that these frame fields would always appear in the Lagrangian in the combination $({\cal E}^{I}{\cal E}^{J}+{\cal F}^{I}{\cal F}^{J})$ and no local scale symmetry is present.

We will refer to the theory (\ref{cec4}) as \emph{Conformal Einstein-Cartan} theory.
Note that if there exist solutions when $e^{I} = \pm f^{I}$ then the first three terms in (\ref{cec4}) become terms familiar from the Einstein-Cartan theory: Palatini, cosmological, and Holst terms respectively \cite{Peldan:1993hi}. The terms proportional to the coefficient $\xi$ vanish in this limit and so represent new behaviour, whilst the final terms quadratic in curvature are boundary terms and do not contribute to the equations of motion.

\subsection{General-Relativistic limit}
\label{grlimit}

By conducting small variations of (\ref{cec4}) one may straightforwardly obtain the equations of motion. Remarkably,  solutions to these equations of motion exist for the ansatz $f^{I} \propto e^{I}$ i.e.

\begin{eqnarray}
f^{I}  = \pm \Omega^{2}(x)e^{I} \label{fome}
\end{eqnarray}
We now show that - as suggested above - such solutions constitute a General-Relativistic limit of the theory. In this limit, we find from combining the $e^{I}$ and $f^{I}$ field equations that 

\begin{equation}
dc=0
\end{equation}
and the field $c$ disappears from the system of field equations. The remaining field equations are equivalent to those obtained from the following action:

\begin{align}
S[\omega,e,\Omega] &= \int \frac{\Omega^{2}}{32\pi \tilde{G}}\left(\pm \epsilon_{IJKL}e^{I}e^{J}R^{KL}-\frac{\Omega^{2}\tilde{\Lambda}}{6}\epsilon_{IJKL}e^{I}e^{J}e^{K}e^{L} \mp \frac{2}{\gamma}R_{IJ}e^{I}e^{J} \right)
\end{align}
Then, varying with respect to $\omega$ and solving for $\omega(e,\Omega)$, eliminating it from the action, we recover the following, second-order action:

\begin{eqnarray}
S[g,\Omega]  =  \int d^{4}x\sqrt{-g} \frac{1}{16\pi G_{0}(\bp)}\left(\Omega^{2}R+6g^{\mu\nu}\partial_{\mu}\Omega\partial_{\nu}\Omega - 2\Lambda_{0}(\bp)\Omega^{4}\right)  \label{conr1}
\end{eqnarray}
where $g_{\mu\nu} \equiv \eta_{IJ}e^{I}_{\mu}e^{J}_{\nu}$, $R$ is the Ricci scalar according to the Christoffel symbols $\Gamma^{\mu}_{\nu\rho}$ and 

\begin{eqnarray*}
\frac{1}{16\pi G_{0}}(\bp) &=&  4\bp (2a_{1}-b_{1}\bp^{2}), \quad \Lambda_{0}(\bp) =  \frac{6(a_{1}-b_{1}\bp^{2}+c_{1}\bp^{4})}{(2a_{1}-\bp^{2}b_{1})}
\end{eqnarray*}
Thus we see in the second-order formalism a kinetic term for $\Omega$ emerges. The definition of $\Omega$ (\ref{fome}) implies that the 
invariance of the action is under the transformation $\{g_{\mu\nu}-\rightarrow e^{2\alpha(x)}g_{\mu\nu},\Omega\rightarrow e^{-\alpha(x)}\Omega\}$ and hence this is the action of a scalar field conformally coupled to gravity. We may utilize the scale gauge freedom in the theory to locally rescale $\Omega$: if we assume that the action is an integration over regions where $\Omega \neq 0$ then a convenient gauge choice is $\Omega = 1$, in which case he action reduces to:

\begin{eqnarray}
S[g]  =  \int  \frac{1}{16\pi G_{0}(\bp)}\left(R- 2\Lambda_{0}(\bp)\right)\sqrt{-g}d^{4}x \label{conr2}
\end{eqnarray}
This is the Einstein-Hilbert action of General Relativity. Therefore the equations of motion obtained in the case $e^{I}\propto f^{I}$ are equivalent\footnote{We note that due to the presence of second-order partial derivatives of the metric tensor in the Einstein-Hilbert action it is necessary to introduce an additional topological term - the {\em Gibbons-Hawking} term - for the variational principle to work out correctly.} to those of General Relativity\footnote{For a different approach to General Relativity and scale invariance see \cite{Gomes:2010fh,Barbour:2011dn,Mercati:2014ama,Anderson:2015ata}.}. Because of this, we refer to the
theory (\ref{conr1}) as \emph{conformalized General Relativity}; if we instead had begun from the Einstein-Hilbert action (\ref{conr2}) we could recover (\ref{conr1}) via a local conformal rescaling of $g_{\mu\nu}$ using $\Omega(x)$.

We must now ask how a theory like General Relativity, with its absolute scales, can emerge from a theory where it is not clear that there is an inbuilt scale. As discussed in the introduction, frequently in the literature an additional `Higgs' scalar $\Phi(x)$ is introduced  alongside $g_{\mu\nu}$ with scales in the gravitational sector to be due to $\Phi(x)$ dynamically reaching constancy; in this context, local scale invariance can be retained via the conformal coupling of $\Phi(x)$ to gravity and this is identical to  the manner in which $\{\Omega,g_{\mu\nu}\}$ combine to yield the locally scale-invariant action (\ref{conr1}). The scalar $\Phi$ would have dimensions of length or mass and would this set a specific scale at each point in spacetime. However, locally in regions where $\Phi(x)\neq 0$, as in the case of the field $\Omega$, we may then readily impose a gauge $\Phi(x)=cst.$ in which the theory would no longer be manifestly scale-invariant. To some this introduction of scale invariance and then its immediate elimination might seem a bit contrived. 

To that end we wish to point out that the breaking of scale invariance in our proposed model is not aided by the introduction by any additional fundamental Higgs fields but is a feature of a specific subclass of solutions including the General-Relativistic solutions characterized by the condition $e^I\propto f^I$. At an extreme, as the equations of motion following from the action (\ref{act5}) are 
polynomial and each term is at least cubic in $\{A^{AB},W^{AB}\}$ then there exist solutions to the unconstrained theory where $\{A^{AB}=0,W^{AB}=0\}$ and the entire $SO(2,4)$-invariance is retained; for such solutions no notion of scale arises from the gravitational sector.
It is conceivable that there may exist solutions to the full theory describing regions where $\{A^{AB}=0,W^{AB}=0\}$ and other regions where these fields are non-vanishing in a particular $SO(2,4)$ gauge and where $e^{I}\propto f^{I}$. In this sense then, the $SU(2,2)\simeq SO(2,4)$ theory may be a candidate for the dynamical origin of scale.

We may characterize the condition $e^{I}\propto f^{I}$ in a different fashion. The General-Relativistic limit is the limit in which a preferred basis $\{U_{(P)}^a,V_{(P)}^a \equiv \epsilon^{a}_{\ph{a}b}U_{(P)}^b,\eta_{ab}U^{a}_{(P)}U^{b}_{(P)}=1\}$ of the $SO(1,1)$ vector space dynamically emerges. This basis is defined by two independent possibilities:

\begin{equation}
U^{a}_{(P)} E^{I}_{\ph{I}a} \overset{*}{=}0  \label{hcon1}
\end{equation}
or
\begin{equation}
V^{a}_{(P)} E^{I}_{\ph{I}a} \overset{*}{=}0  \label{hcon2}
\end{equation}
where recall the definition of $E^{Ia}$ from (\ref{aab2}). Meanwhile, the ansatz (\ref{eanz}) can be written equivalently as follows:

\begin{eqnarray}
E^{Ia} & \overset{*}{=} &  \frac{1}{2}(e^{I}+f^{I})U^{a}+  \frac{1}{2}(e^{I}-f^{I})V^{a} \label{eia}
\end{eqnarray}
where $\eta_{ab}U^{a}U^{b} = 1$ and $V^{a} \equiv \epsilon^{a}_{\ph{a}b}U^{b}$; here $U^a$ and $V^a$ should be regarded as an arbitrary choice of basis of the two-dimensional vector space. For example in the case of the condition (\ref{hcon1}) holding, we have - using (\ref{eia}) and (\ref{hcon1}) -  that

\begin{eqnarray}
f^{I}  &=&  - \frac{\left(U^{a}U_{(P)a} + V^{a}U_{(P)a}\right)}{\left(U^{a}U_{(P)a}-V^{a}U_{(P)a}\right)} e^{I} \equiv - \Omega^{2} e^{I} \label{eomegaf}
\end{eqnarray}
Thus, if we choose the preferred gauge $U^{a} = U_{(P)}^{a}$ as our basis then in this gauge we have from (\ref{eomegaf}) that $f^{I} \overset{*}{=} -e^{I}$. If, alternatively, the condition (\ref{hcon2}) holds then $f^{I} = +\Omega^{2} e^{I}$ and in the preferred gauge $f^{I} \overset{*}{=} e^{I}$. Choosing the preferred $SO(1,1)$ gauge corresponds to choosing the scale gauge in which $|\Omega| = 1$.

In some respects this is similar to the breaking of rotational invariance for a ferromagnet: although a preferred direction $\overrightarrow{X}$ appears as a property of low-temperature solutions this direction $\overrightarrow{X}$ makes no appearance in the fundamental equations of motion. Similarly, a preferred direction in the $SO(1,1)$ space - $U_{(P)}^{a}$- appears for General-Relativistic solutions but its existence is a property of some solutions rather than a basic constituent of the theory. Neither can such a preferred vector $U_{(P)}^a$ be defined uniquely outside the General-Relativistic limit.

\subsection{Relation to Weyl and fourth-order conformal gravity}
\label{weylrel}
In the previous section we considered a theory recovered by `freezing' all the degrees of freedom in $W^{AB}$ to take a specific form; the resulting theory possessed a local $SO(1,3)$ invariance as well as independent rescaling invariances under $e^{I}\rightarrow e^{\alpha(x)}e^{I}$, $f^{I}\rightarrow e^{-\alpha(x)}f^{I}$, $c\rightarrow c - d\beta(x)$. An interesting property of this theory is that  beginning from the field equations for the set $\{e^{I},f^{I},\omega^{I}_{\ph{I}J},c\}$ we may solve algebraically for \emph{any} one of $\{e^{I},f^{I},\omega^{I}_{\ph{I}J}\}$ and eliminate it from the action. In Appendix \ref{solvf}, this is explicitly illustrated for the following action:

\begin{align}
S[\omega,c,e,f]=\int \alpha \epsilon_{IJKL}e^If^JR^{KL}+\beta \epsilon_{IJKL}e^Ie^Jf^Kf^L+\gamma e^If_I dc \label{hhem}
\end{align}
where $\alpha,\beta,\gamma$ are constants. Solving algebraically for the field $f^{I}$ from its own equation of motion and inserting this solution back into the action, one obtains:

\begin{align}
S[\omega,c,e] &= \frac{1}{\beta}\int \frac{\alpha^{2}}{4}\epsilon_{IJKL}C^{IJ}C^{KL} -6\gamma^2 dc*dc +11\alpha\gamma e_KR^K dc-\frac{\alpha^{2}}{4}\epsilon_{IJKL}R^{IJ}R^{KL } \label{conm}
\end{align}
where $*$ is the Hodge dual operator built from the field $e^{I}$, $R^{J}\equiv  ((e^{-1})_{I}\lrcorner R^{IJ}) $ is the Ricci one-form and 

\begin{eqnarray}
C^{IJ} &\equiv & R^{IJ} -  6e^{I}\left(R^{J}-\frac{R}{6}e^{J}\right)
\end{eqnarray}
where $R$ is the Ricci scalar. We now consider the effect of placing an additional constraint on this theory.
Recall that in the Einstein-Cartan theory, the equation $D^{(\omega)}e^{I} =0 $ was the equation of motion for $\omega^{I}_{\ph{I}J}$ and allowed one to solve for $\omega^{I}_{\ph{I}J}$ and eliminate it from the action principle. Now consider the following generalisation of this equation:

\begin{eqnarray}
D^{(\omega+c)}e^{I}  &=&  de^{I} + \omega^{I}_{\ph{I}J}e^{J} + c e^{I} = 0 \label{domega}
\end{eqnarray}
This equation is invariant under a more restricted group of symmetry transformations of the Conformal Einstein-Cartan theory: that when $\alpha(x)=\beta(x)$. This equation is \emph{not} the equation of motion for $\omega^{I}_{\ph{I}J}$ one would get by varying with respect to $\omega^{I}_{\ph{I}J}$ for this theory but we may enforce it via a Lagrangian constraint. 
Doing so, we may now use (\ref{domega}) to solve for $\omega^{I}_{\ph{I}J}(e,c)$ and substitute this solution into the action (\ref{conm}). The resulting action is a functional only of $e^{I}$ (appearing via $\{e_{I},(e^{-1})^{I\mu},R^{IJ}(e)\}$) and $c$ (via $dc$):

\begin{eqnarray}
S[e,c]&=& \frac{1}{\beta}\int \frac{\alpha^{2}}{4}\epsilon_{IJKL}{\cal C}^{IJ}{\cal C}^{KL} -6\gamma^2 dc*dc +11\alpha\gamma e_KR^K dc-\frac{\alpha^{2}}{4}\epsilon_{IJKL}R^{IJ}R^{KL }
\end{eqnarray}
This is the action for \emph{fourth-order Weyl gravity} which, as the name suggests, yields field equations containing fourth-derivatives of fields. If we further constrain $dc=0$ then we recover the action:

\begin{align}
S[e] &= \frac{1}{\beta}\int \frac{\alpha^{2}}{4}\epsilon_{IJKL}C^{IJ}(e)C^{KL}(e)-\frac{\alpha^{2}}{4}\epsilon_{IJKL}R^{IJ}(e)R^{KL }(e) \label{conm2}
\end{align}
The quantity ${\cal C}^{IJ}(e)$ is the Weyl two-form, related to the Weyl tensor ${\cal C}_{\mu\nu\alpha\beta} = {\cal C}_{\mu\nu}^{\ph{\mu\nu}IJ}e_{I \alpha}e_{J \beta}$ via:

\begin{eqnarray}
{\cal C}^{IJ}  =  \frac{1}{2}{\cal C}_{\mu\nu}^{\ph{\mu\nu}IJ} dx^{\mu}dx^{\nu}
\end{eqnarray}
The action (\ref{conm2}) is thus proportional to that of \emph{fourth-order conformal gravity} plus a boundary term quadratic in $R^{IJ}$. Hence, fourth-order conformal gravity can be recovered from the original $SU(2,2)$ gauge theory via the implementation of a number of constraints. Indeed, this was the result found 
by Kaku, Townsend, and Van Nieuwenhuizen \cite{Kaku:1977pa}. Their approach was essentially the same as the steps discussed in this section  i.e. fourth-order conformal gravity was recovered by implicitly constraining the symmetry breaking fields that break $SO(2,4)\rightarrow SO(1,3)\times SO(1,1)$ and explicitly constraining the spin-connection $\omega^{I}_{\ph{I}J}$. For the particular action they considered (specifically the action (\ref{hhem}) with $\gamma=0$) $dc$ vanishes automatically from the action and did not need to be constrained to vanish. The relation of this approach of recovering Weyl gravity from a gauge theory of gravity to Cartan's conception of geometry has recently been discussed in detail\footnote{The link between Cartan geometry and conformal physics has previously been investigated in the case of $2+1$ spacetime dimensions \cite{Gryb:2012qt}.}  \cite{Attard:2015jfr}.

It has become somewhat common lore that fourth-order Weyl gravity is the gauge theory of the conformal group \cite{Hassan:2013pca,Joung:2014aba} . However, if one is looking to cast gravity as a gauge theory akin to those of particle physics, why completely freeze all the degrees of freedom in the symmetry breaking fields? The analogue in electroweak theory would be an insistence that $\varphi^{\dagger}\varphi$ for the electroweak Higgs $\varphi$ were fixed to be a constant - this would force a non-vanishing expectation value for $\varphi$ much as a non-vanishing expectation value for $W^{AB}$ was achieved in the above approach. The discovery of the Higgs boson demonstrates that in that case it would be incorrect to apply such constraints;
should gravity be any different? Even allowing this, why then further constrain $\omega^{I}_{\ph{I}J}$ to take a solution that would not generally follow from $\omega^{I}_{\ph{I}J}$'s equation of motion \cite{Wheeler:2013ora}?

\section{Vacuum solutions of the full theory}
\label{maxsym}
We now `un-freeze' the field $W^{AB}$. Its classical evolution will now be dictated entirely by
its own equations of motion in conjunction with those of other fields. We will demonstrate that there exist simple solutions to the theory in which 
the field $W^{AB}$ has a non-vanishing, constant expectation value and one may interpret the accompanying spacetime geometry as being de Sitter or anti de Sitter space. We will look for solutions where $W^{AB}$ takes the following form:

 \begin{equation}
 W^{AB} \overset{*}{=} \left( \begin{array}{cc}
0 & 0 \\
0 & \phi \epsilon^{ab}
 \end{array} \right) \label{wtolorentz}
 \end{equation}
where recall that $a,b,c,\dots$ are $SO(1,1)$ indices and the gauge-fixing condition $W^{Ia} \overset{*}{=}0$ has been imposed. We will focus on searching for solutions where

\begin{eqnarray}
\phi =\bp = \mathrm{cst.}
\end{eqnarray}
As detailed in the previous section, this form of $W^{AB}$ (even if $\phi$ were not constant) breaks the original $SO(2,4)$ symmetry of the theory down to $SO(1,3)\times SO(1,1)$ if the signature of $\eta_{ab}$ is $(-,+)$; we will assume this to be the case. Clearly, the existence of solutions satisfying this condition does not indicate that they are dynamically favoured. In this paper our analysis will be limited to establishing linear stability of them with respect to small perturbations.

Now we turn to the form of the connection $A^{AB}$. Given the above symmetry breaking, a general ansatz for this field is:

 \begin{align}
 A^{AB} &\overset{*}{=} \left( \begin{array}{cc}
\omega^{IJ} & E^{Ia} \\
-E^{Ia} & c \epsilon^{ab}
 \end{array} \right) \label{ahab}
 \end{align}
For the remainder of this section we will remain in the gauge implied by the form (\ref{wtolorentz}) and so, subsequently, $=$ should be taken to mean an equality that holds given this gauge condition. It follows from (\ref{wtolorentz}) and (\ref{ahab}) that

 \begin{align}
 DW^{AB} &= \left( \begin{array}{cc}
0 & \phi E^{I}_{\ph{I}c}\epsilon^{ca} \\
-\phi E^{I}_{\ph{I}c}\epsilon^{ca} & 0
 \end{array} \right)
 \end{align}
and we make an ansatz for $E^{Ia}$ such that :
 \begin{align}
DW^{AB} &= \left( \begin{array}{cc}
0 & e^{I}u^{a} \\
-e^{I}u^{a} & 0 
 \end{array} \right) \label{80sstallone}
 \end{align}
Note that as in Section \ref{grlimit}, we are effectively restricting ourselves to the case where $E^{Ia}$ contains a single independent frame field; this seems a reasonable assumption given that we are searching for solutions interpretable in terms of a single spacetime metric. The form of the ansatz (\ref{80sstallone}) is invariant under a local rescaling $u^{a}\rightarrow e^{\alpha(x)}u^{a}$, $e^{I}\rightarrow e^{-\alpha(x)}e^{I}$. The $SO(2,4)$ curvature two-form then takes the following form:
 
 \begin{align}
 F^{AB} &= \left( \begin{array}{cc}
 R^{IJ}+\frac{u^{2}}{\phi^{2}}e^{I}e^{J} &  -\frac{1}{\bar{\phi}}\left(D^{(c)}v^{a}e^{I}+v^{a}D^{(\omega)}e^{I}\right) \\
-F^{Ia} & dc \epsilon^{ab}
 \end{array} \right)
 \end{align}
where $v^{a}\equiv \epsilon^{a}_{\ph{a}b}u^{b}$ and $D^{(c)}$ and $D^{(\omega)}$ are the $SO(1,1)$ and $SO(1,3)$ covariant derivatives respectively. We look for solutions where 

\begin{eqnarray}
F^{Ia}=0 \label{fia}
\end{eqnarray}
and it can be shown that this implies that:

\begin{eqnarray}
\frac{1}{2u^{2}} du^{2} e^{I}+D^{(\omega)}e^{I} &\equiv & \frac{1}{2u^{2}} du^{2} e^{I}+de^{I}+\omega^{I}_{\ph{I}J}e^{J}  = 0 \label{cond1} \\
u_{a}\epsilon^{ab}D^{(c)}u_{b} &=&  u_{a}\epsilon^{a}_{\ph{a}b}du^{b} + u^{2}c =  0 \label{cond2}
\end{eqnarray}
From equations (\ref{cond1}) and (\ref{cond2}) we then have that 

\begin{align}
\omega_{\mu}^{\ph{\mu}IJ} &= 2 e^{\nu [I}\partial_{[\mu}e_{\nu]}^{J]}+ e_{\mu K} e^{\nu I} e^{\alpha J}\partial_{[\alpha}e_{\nu]}^{K} \nn\\
&  +  \frac{1}{u^{2}}e^{\nu [I}e_{[\nu}^{J]}\partial_{\mu]}u^{2}+ \frac{1}{2u^{2}}e_{\mu K} e^{\nu I} e^{\alpha J}e^{K}_{[\nu}\partial_{\alpha]}u^{2} \label{omegae}\\
c_{\mu} &=   -\frac{1}{u^{2}}u_{a}\epsilon^{a}_{\ph{a}b}\partial_{\mu}u^{b} \label{cond3}
\end{align}
We note that these expressions are invariant under the simultaneous local rescaling $u^{a}\rightarrow e^{\alpha(x)}u^{a}$ and $e^{I}\rightarrow e^{-\alpha(x)}e^{I}$. From (\ref{cond3}) we have that:

\begin{eqnarray}
dc  &=&   - d\left(\frac{1}{u^{2}}\right)u_{a}\epsilon^{a}_{\ph{a}b}du^{b} - \frac{1}{u^{2}} \epsilon_{ab} du^{a}du^{b} -\frac{1}{u^{2}}u_{a}\epsilon^{a}_{\ph{a}b} ddu^{b} \label{dcee}
\end{eqnarray}
If $u^{2}\neq 0$ and of consistent sign throughout the spacetime we are considering, then we may choose an $SO(1,1)$ gauge where $u^{a}= \sqrt{u^{2}} \delta^{a}_{4}$ (for $u^{2}=1$) or $u^{a}=\sqrt{-u^{2}}\delta^{a}_{-1}$ (for $u^{2}=-1$). One can see that in this gauge (and hence all $SO(1,1)$ gauges) the first and second terms in (\ref{dcee}) disappear whilst the third term disappears due to the identity exterior derivative identity $dd=0$.
Thus $dc=0$ for our ansatz.
Furthermore we propose that the $SO(1,3)$ curvature takes the following form:

\begin{eqnarray}
R^{IJ}  =  \lambda e^{I}e^{J} \label{curv}
\end{eqnarray}
where $\lambda$ is a constant. We will assume that $u^{2}\neq 0$ and of consistent sign and for convenience we will
utilize the freedom to locally rescale $u^{a}$ by setting $u^{2} = cst.$.
Then,  given the expression $\omega^{I}_{\ph{I}J}(e)$ of (\ref{omegae}), the ansatz (\ref{curv}) implies that the solution for $g_{\mu\nu} \equiv \eta_{IJ}e^{I}_{\mu}e^{J}_{\nu}$ will be the metric de Sitter space ($\lambda >0$) or anti-de Sitter space ($\lambda <0$). Therefore in summary the $SO(2,4)$ curvature is assumed to \emph{on-shell} take the following simple form:
\begin{equation}
F^{AB} = \left( \begin{array}{cc}
\left(\lambda+\frac{u^{2}}{\phi^{2}}\right)e^{I}e^{J} & 0 \\
0 & 0
 \end{array} \right)
 \end{equation}
From the $A^{Jb}$ equations of motion we find that only $\{a_{1},b_{1},c_{1}\}$ terms of (\ref{act5}) offer a non-vanishing contribution and that these equations of motion provide a value for the cosmological constant $\lambda$ in terms of parameters $\{a_{1},b_{1},c_{1}\}$ and $\bp^{2}$: 

\begin{eqnarray}
\lambda  &=&  -\frac{u^{2}}{\bp^{2}}\frac{\left(a_{1}-b_{1}\bp^{2}+c_{1}\bp^{4}\right)}{\left(a_{1}-\frac{\bp^{2}b_{1}}{2}\right)} \label{lamda}
\end{eqnarray}
A solution for the value of $\bp$ may be obtained by looking at the equations of motion obtained by varying $W^{ab}$. Again, only $\{a_{1},b_{1},c_{1}\}$ terms yield a non-vanishing contribution, and the equation reads:

\begin{eqnarray}
%\epsilon_{IJKL}g^{I}g^{J}g^{K}g^{L}\frac{u^{4}}{\bp^{2}}\left(2a_{1}\xi^{2}-6b_{1}\xi\bp^{2}+
%10c_{1}\bp^{4}\right)  = 0
\left(2a_{1}\xi^{2}-6b_{1}\xi\bp^{2}+
10c_{1}\bp^{4}\right)u^{4}  = 0 \label{scal}
\end{eqnarray}
where $\xi \equiv 1+ \lambda \bp^{2}/u^{2}$. If we now assume that $u^{2}\neq 0$, equations (\ref{lamda}) and (\ref{scal}) may be combined to obtain the following equation:

\begin{eqnarray}
0 &=&  \bp\frac{\left(b_{1}^{2}-4a_{1}c_{1}\right)\left(-5a_{1}+3b_{1}\bp^{2}-c_{1}\bp^{4}\right)
}{(b_{1}\bp^{2}-2a_{1})} \label{eom1}
\end{eqnarray} 
This equation may be seen as a defining equation for $\bp^{2}$ -assuming that the special case $b_{1}^{2}-4a_{1}c_{1} =0$ does not apply- i.e. we may solve it to find $\bp^{2}(a_{1},b_{1},c_{1})$. This restricts the $\{a_{1},b_{1},c_{1}\}$ parameter space to values where real, positive solutions for $\bp^{2}$ exist. If we use this result in (\ref{lamda}) we recover the simple relation

\begin{eqnarray}
\lambda  &=&   \frac{4u^{2}}{\bp^{2}}
\end{eqnarray}
Thus, $u^{2}$ determines the sign of the cosmological constant $\lambda$.
We see then that extremely simple solutions to the theory exist, just as in the case of General Relativity, and that many of the degrees of freedom of the theory do not contribute within these solutions (e.g. it is assumed that $W^{IJ}=0$). 

An important point is that the existence of a constant solution for the field $\phi$ depended on our ansatz $R^{IJ} = \lambda e^{I}e^{J}$. This is because the $(a,b)$ component of the $W^{AB}$ equation of motion contains a term proportional to $a_{1}\epsilon_{IJKL}R^{IJ}R^{KL}$; only when $R^{IJ} = \lambda e^{I}e^{J}$ does this term appear on the same footing as terms structurally similar to potential terms i.e. for a potential $\chi(\phi)$, contributions of the form $(d\chi/d\phi)\epsilon_{IJKL}e^{I}e^{J}e^{K}e^{L}$. If $R^{IJ}$ is of a more general form then static solutions for $\phi$ likely will not always exist. This is highly reminiscent of the case of models of gravity based on the groups $SO(1,4)|SO(2,3)$ wherein it was rather more common for $\phi$ to `roll' down an effective potential rather than be static \cite{Magueijo:2013yya}. 

\section{Perturbations}
\label{perts}
An important property of General Relativity is the stability of physically interesting solutions such as Minkowski space or de Sitter space with respect to small perturbations. We will now see whether the maximally symmetric solutions to the $SU(2,2)$ model of gravity are linearly stable. To do this, we will take the maximally symmetric solutions to constitute a background spacetime and and consider small perturbations to the background form that the fields $\{A^{AB},W^{AB}\}$ take. We will then expand the Lagrangian for the theory (incorporating all terms) around the background solutions up to quadratic order in smallness. This will tell us whether these background spacetimes are linearly stable or not.

At the level of perturbations we can retain the gauge fixing $W^{Ia}\overset{*}{=} 0$ and we will use bars above quantities to denote that they are background quantities (e.g. $\bar{A}^{AB}$ denotes the background form of $A^{AB}$). As for the case of the background solutions, we fix the local rescaling symmetry amongst the background pair $\{\bar{u}^{a}, \bar{e}^{I}\}$ so that $|\bar{u}^{2}| = 1$.
A completely general ansatz for small perturbations is as follows:

 \begin{equation} \delta W^{AB} =\bp \left( \begin{array}{cc}
\delta H^{IJ} & 0 \\
0 & \delta\alpha \epsilon^{ab}
 \end{array} \right) , \quad \delta A^{AB} = \left( \begin{array}{cc}
\delta \omega^{IJ} & -\frac{\bar{v}^{a}}{\bar{\phi}}\delta e^{I} - \frac{\bar{u}^{a}}{\bar{\phi}}\delta h^{I} \\
-\delta A^{Ia} & \delta c \epsilon^{ab}
 \end{array} \right)
 \end{equation}
We may insert this ansatz into the Lagrangian four-form for the theory to obtain a Lagrangian quadratic in smallness. The perturbed total Lagrangian $\delta L$ decomposes into two independent parts: a part that depends on the variables $\{\delta e^{I},\delta \omega^{IJ},\delta\alpha\}$ and a part that depends on the variables $\{\delta h^{I},\delta c,\delta H^{IJ}\}$:

\begin{equation}
\delta L = \delta L_{\{0,2\}}(\delta e,\delta \omega,\delta\alpha) + \delta L_{1}(\delta h,\delta c,\delta H)
\end{equation}
As will be seen, the labels $\{0,2\}$ and $1$ refer to the spin of the perturbations from each part of the Lagrangian when cast into a second-order form. We first concentrate on $\delta L_{\{0,2\}}$. It is extremely useful to make a variable redefinition to help cast terms in a more familiar way:
 after calculation, we find amongst the terms the following contribution to the perturbed Lagrangian:

\begin{eqnarray}
\epsilon_{IJKL}\bar{e}^{I}\db \delta \omega^{KL}\left(2\bar{u}^{2}\bar{e}^{J}\delta \alpha \left(b_{1}\bp-2\left(\frac{\lambda\bp^{2}}{\bar{u}^{2}}-1\right) \frac{a_{1}}{\bp}\right)  + 4\bar{u}^{2}\left(b_{1}\bp-\frac{2a_{1}}{\bp}\right)(\bar{e}^{I}\delta\alpha+\delta e^{I})\right) \label{mess}
\end{eqnarray}
If we define the following variables:

\begin{align}
\bar{\theta}^{I}  &=   \sqrt{2\bar{u}^{2}(b_{1}\bp-2\frac{a_{1}}{\bp})}\bar{e}^{I}\\
\delta \theta^{I} &=  \sqrt{2\bar{u}^{2}(b_{1}\bp-2\frac{a_{1}}{\bp})} \delta e^{I} + \left(\frac{d\left( \sqrt{2\bar{u}^{2}(b_{1}\bp-2\frac{a_{1}}{\bp})}\bp\right)}{d\bp} -2\frac{a_{1}}{ \sqrt{2\bar{u}^{2}(b_{1}\bp-2\frac{a_{1}}{\bp})}}\lambda \bp \right) \bar{e}^{I}\delta\alpha 
\end{align}
then the terms (\ref{mess}) become:

\begin{eqnarray}
2\epsilon_{IJKL}\bar{\theta}^{I}\delta \theta^{J}\db \delta \omega^{KL}+ \epsilon_{IJKL}\bar{\theta}^{I}\bar{\theta}^{J}\delta \omega^{K}_{\ph{I}M}\delta \omega^{ML}
\end{eqnarray}
This is simply equivalent to the perturbation to the Einstein-Palatini action $\epsilon_{IJKL}e^{I}e^{J}R^{KL}(\omega)$ involving $\delta\omega^{IJ}$ around a background $\theta^{I} = \bar{\theta}^{I}$. All dependency upon $\delta\alpha$ has disappeared.
We can make further progress by decomposing the spin connection perturbation $\delta\omega^{IJ}$ into a `torsion-free' part $\delta\tilde{\omega}^{IJ}$ and the contorsion $\delta C^{IJ}$:

\begin{eqnarray}
\delta \omega^{IJ} &=& \delta\tilde{\omega}^{IJ} + \delta C^{IJ}
\end{eqnarray}
where $\delta\tilde{\omega}^{IJ}(\delta \theta,\partial \delta \theta)$ is the solution to the equation:

\begin{eqnarray}
d\delta \theta^{I}  + \bar{\omega}^{I}_{\ph{I}J}\delta \theta^{J}  + \delta\tilde{\omega}^{I}_{\ph{I}J}\bar{\theta}^{J} &=& 0
\end{eqnarray}
Calculations then suggest that 
\begin{align}
\delta L_{\{0,2\}}  &=  2\epsilon_{IJKL}\bar{\theta}^{I}\delta \theta^{J}\db \delta \tilde{\omega}^{KL}(\theta)+ \epsilon_{IJKL}\bar{\theta}^{I}\bar{\theta}^{J}\delta \tilde{\omega}^{K}_{\ph{I}M}(\theta)\delta \tilde{\omega}^{ML}(\theta)\nn\\
&-4\chi \epsilon_{IJKL}\delta \theta^{I}\delta \theta^{J} \bar{\theta}^{K}\bar{\theta}^{L}+\epsilon_{IJKL}\bar{\theta}^{I}\bar{\theta}^{J}\delta C^{K}_{\ph{I}M}\delta C^{ML}\nn \\
&+\bar{u}^{2}\frac{a_{2}}{\mu^{2}}\left(\delta C^{IK}\delta C_{K\ph{J}}^{\ph{K}J}-2\left(\frac{1}{\mu}\frac{d\mu}{d\bp}\bp  -4a_{1}\bp \chi +1\right) d\delta \alpha \delta C^{IJ}\right)\bar{\theta}_{I}\bar{\theta}_{J} \nn\\
&
 -\left(\frac{1}{2}\frac{d^{2}\chi}{d\bp^{2}}-64\chi^{3}a_{1}^{2}\right)\bp^{2}(\da)^{2}\epsilon_{IJKL}\bar{\theta}^{I}\bar{\theta}^{J}\bar{\theta}^{K}\bar{\theta}^{L}
\end{align}
where

\begin{eqnarray}
\mu &\equiv&  \sqrt{2\bar{u}^{2}(b_{1}\bp-2\frac{a_{1}}{\bp})}, \quad \chi \equiv  \frac{1}{2}\frac{\left(\frac{a_{1}}{\bp^{3}}-\frac{b_{1}}{\bp}+c_{1}\bp\right)}{\left(b_{1}\bp-2\frac{a_{1}}{\bp}\right)^{2}}
\end{eqnarray}
We may now vary with respect to $\delta C^{IJ}$; we find that the resulting equation of motion is an algebraic equation with which we can actually solve for $\delta C^{IJ}$ in terms of  $\partial \delta \alpha$. This solution may be re-inserted into the Lagrangian to eliminate $\delta C^{IJ}$ from the action, yielding:
\begin{align}
\delta L_{\{0,2\}} &=  \delta L_{2} + \delta L_{0}  \\
\delta L_{2}(\delta e)&= 2\epsilon_{IJKL}\bar{\theta}^{I}\delta \theta^{J}\db \delta \tilde{\omega}^{KL}(\theta)+ \epsilon_{IJKL}\bar{\theta}^{I}\bar{\theta}^{J}\delta \tilde{\omega}^{K}_{\ph{I}M}(\theta)\delta \tilde{\omega}^{ML}(\theta)\nn\\
& -4\chi \epsilon_{IJKL}\delta \theta^{I}\delta \theta^{J} \bar{\theta}^{K}\bar{\theta}^{L} \\
\delta L_{0}(\delta \alpha) &=-\left(\frac{a_{2}^{2}}{8\mu^{4}+2a_{2}^{2}}\right) \left(\frac{1}{\mu}\frac{d\mu}{d\bp}\bp  -4a_{1}\bp \chi +1\right)^{2}\left(\partial_{M}\delta\alpha\partial^{M}\delta\alpha \right)\epsilon_{IJKL}\bar{\theta}^{I}\bar{\theta}^{J}\bar{\theta}^{K}\bar{\theta}^{L}\nn \\
&
 -\left(\frac{1}{2}\frac{d^{2}\chi}{d\bp^{2}}-64\chi^{3}a_{1}^{2}\right)\bp^{2}(\da)^{2}\epsilon_{IJKL}\bar{\theta}^{I}\bar{\theta}^{J}\bar{\theta}^{K}\bar{\theta}^{L} 
 \end{align}
We see then that this part of the perturbed Lagrangian decomposes into two independent pieces. The first piece, $\delta L_{2}$, depends solely on $\delta \theta$; it is equivalent 
to the perturbed Lagrangian of General Relativity against a background maximally symmetric spacetime with co-tetrad $\bar{\theta}^{I}$ and cosmological constant
$\Lambda = 12\chi$. Therefore, perturbations include spin-2 gravitational wave solutions, as in General Relativity.

The second piece, $\delta L_{0}$ depends solely on perturbations $\delta\alpha$. The first term is a correct-sign kinetic term for $\delta\alpha$, whereas the second term acts as a mass term for the field, as one would expect for perturbations around a solution $\phi = \mathrm{cst.}$. However, note that the term is not due solely to the `second derivative of a potential' (here $d^{2}\chi/d\bp^{2}$) term as one might expect in a theory of a scalar field in curved spacetime - there is an additional term due to the $a_{1}$ in the action; this reflects the direct coupling between $\phi$ and a term quadratic in spacetime curvature (this corresponds to the ${\cal C}_{1}$ term of (\ref{cec4}) with $\phi$ `un-frozen'). The sign of this mass-squared term is not of definite sign for all $\{a_{1},b_{1},c_{1}\}$; its sign is given by:

\begin{eqnarray}
&&\mathrm{sign}\left(\frac{d^{2}\chi}{d\bp^{2}}-128\chi^{3}a_{1}^{2} = \frac{40a_{1}b_{1}\bp^{2}-100a_{1}^{2}-3b_{1}^{2}\bp^{4}}{(b_{1}\bp^{3}-2a_{1}\bp)^{3}}\right)
\end{eqnarray}
where recall that the background value $\bp$ is a function of $\{a_{1},b_{1},c_{1}\}$. Hence, positivity
of effective mass-squared of the new propagating spin-0 degree of freedom places a restriction upon the parameter space $\{a_{1},b_{1},c_{1}\}$.
Recall our earlier field redefinition $\bar{\theta}^{I} = \mu \bar{e}^{I}$; this implies that we must have that $\mu \equiv  \sqrt{2\bar{u}^{2}(b_{1}\bp-2\frac{a_{1}}{\bp})}$ is real and so

\begin{eqnarray}
2\bar{u}^{2}\bp\left(b_{1}-2\frac{a_{1}}{\bp^{2}}\right) > 0
\end{eqnarray}
This implies that the sign of the mass-squared is

\begin{eqnarray}
 \mathrm{sign} \left(\bar{u}^{2}\left(40a_{1}b_{1}\bp^{2}-100a_{1}^{2}-3b_{1}^{2}\bp^{4}\right)\right) \label{massign}
\end{eqnarray}
The sign of the cosmological constant in the background is given by $\mathrm{sign}(\bar{u}^{2})$ and so if we take a positive cosmological constant ($\bar{u}^{2}>0$)  then (\ref{massign}) implies that oscillations of the scalar field $\delta\alpha$ are only stable if the combination of $\{a_{1},b_{1},\bp\}$ in parenthesis is positive:

\begin{eqnarray}
\left(40a_{1}b_{1}\bp^{2}-100a_{1}^{2}-3b_{1}^{2}\bp^{4} \right)> 0
\end{eqnarray}
We can immediately see that if $a_{1}=0$ then the degree of freedom $\delta \alpha$ is not stable: there would exist exponentially growing solutions for the field $\delta\alpha$ which implies that the background solution would not be stable. Therefore the presence of $a_{1}$ and $b_{1}$ terms together can ensure positivity of the effective mass-squared of $\delta\alpha$, yielding small oscillations of $\delta\alpha$ as solutions to its perturbative equation of motion.

We now look at the remaining part of the perturbed Lagrangian: $\delta L_{1}$. Making use of the background relation $\lambda =4\bar{u}^{2}/\bp^{2}$, this part of the Lagrangian simplifies to:

\begin{align}
\delta L_{1}(\delta H,\delta h,\delta c)  &=  4\left(\ast \delta H-\xi_{2}\delta N\right) d\delta c
  +6\xi_{2}\delta N\delta N+2\xi_{2} d\delta c d\delta c +\frac{5}{4}\xi_{4}\delta H \ast \delta H -\xi_{4}\nu \delta N\ast \delta H\nn \\
&  +\xi_{3}\left(-2 d\delta c+2\delta N + \frac{10}{\nu}\delta H\right)\left(-2 d\delta c+2\delta N + \frac{10}{\nu}\delta H\right)
 \end{align}
 where

\begin{eqnarray}
\quad \delta H &\equiv &  \frac{1}{2}\delta H_{MN}\bar{\theta}^{M}\bar{\theta}^{N}, \quad \delta N =   \frac{\bar{u}^{2}}{ \mu}\bar{\theta}_{I}\delta h^{I}
 \nn\\
\xi_{2} &\equiv& \bp^{2} a_{2}, \quad  \xi_{3}  \equiv   \bp^{2} a_{3} ,\quad \xi_{4} \equiv  \frac{32}{\mu^{4}\bp} (5b_{1}-2c_{1}\bp^{2}), \quad \nu \equiv \frac{\mu^{2}\bp^{2} }{\bar{u}^{2}}\nn
\end{eqnarray}
Furthermore we have defined the Hodge star/Hodge dual operator on a two-form $T$ relative to the background frame field $\bar{\theta}^{I}$ as:

\begin{eqnarray}
 \ast T \equiv    \frac{1}{4}\epsilon_{IJKL}T^{IJ}\bar{\theta}^{K}\bar{\theta}^{L}
 \end{eqnarray}
To clean up notation we have defined the two-forms $\delta H$ (which comes from $\delta W^{AB}$) and $\delta N$ (which comes from $\delta A^{AB}$). These fields appear algebraically in the action and we may actually solve for each of them in terms of $d\delta c$ and $\ast d\delta c$. We first note a perhaps surprising structural feature of $\delta L_{1}$. If the perturbation to $\delta W^{AB}$ is switched-off then we recover:

\begin{align}
\delta L_{1}(\delta H = 0,\delta h,\delta c)  &=  -4\xi_{2}\delta N d\delta c
  +6\xi_{2}\delta N\delta N+2\xi_{2} d\delta c d\delta c\nn\\
  &  +\xi_{3}\left(-2 d\delta c+2\delta N \right)\left(-2 d\delta c+2\delta N \right)
 \end{align}
The resulting equation of motion for $\delta N$ reveals that $\delta N$ is simply proportional to $\delta c$. Using this solution back in $\delta L_{1}$ we recover:

\begin{eqnarray}
\delta L_{1}(\delta H = 0,\delta c)  &\propto & d\delta c d\delta c
 \end{eqnarray}
This is a boundary term (the equation of motion obtained by varying $\delta c$ is the identity $dd\delta c =0$) and so we see that in the absence of the perturbation to the Higgs field $W^{AB}$, there is no corresponding second-order dynamics for $\delta c$! Allowing for a non-zero $\delta H$, the end result of eliminating $\delta N$ and $\delta H$ is rather complicated, but is of the general form:
 
 \begin{eqnarray}
 \delta L_{1}   &=&   - f_{1}(\xi_{i},\nu,\bp) d\delta c \ast d\delta c + f_{2}(\xi_{i},\nu,\bp)d\delta c d\delta c
\end{eqnarray}
 We see the first term is a Maxwell-type kinetic term for the field $\delta c$. The `right sign' of such a term is $-d\delta c \ast d\delta c$. 
 
 \begin{eqnarray}
\delta L_{1}  &=&  -\frac{20}{\xi_{4}}\left(\frac{(-12\xi_{2}+\xi_{2}\xi_{4}\nu)^{2}}{(30\xi_{2})^{2}+(\xi_{4}\nu^{2})^{2}}\right)d\delta c\ast d\delta c + f_{2}(\xi_{i},\nu,\bp)d\delta cd\delta c
 \end{eqnarray}
 and in the limit $\xi_{2} \rightarrow 0$:
 
  \begin{eqnarray}
\delta L_{1}  &=&  -\frac{20^{2}}{\xi_{4}}\left(\frac{(-4\xi_{3}+\xi_{4}\xi_{3}\nu)^{2}}{(100 \xi_{3})^{2}+(\xi_{4}\nu^{2})^{2}}\right)d\delta c\ast d\delta c + f_{2}(\xi_{i},\nu,\bp)d\delta cd\delta c
 \end{eqnarray}
In both limits, the sign of the term in front of $d\delta c \ast d \delta c$ is given by the sign of $\xi_{4}$, and so we require in these limits that:

\begin{equation}
\frac{(5b_{1}-2c_{1}\bp^{2})}{\bp} > 0
\end{equation}
In conclusion then, the spectrum of perturbations around maximally symmetric background solutions is that of General Relativity (via the field $\delta \theta^{I}$), a massive scalar field (via $\delta \alpha$), and a massless one-form field (via $\delta c$). Therefore, around such backgrounds, the degrees of freedom of the theory are those of a \emph{tensor vector scalar} theory; variants of such theories have been explored as alternative theories of gravity \cite{Bekenstein:2004ne, Moffat:2011rp} though the $SU(2,2)$ theory differs from these examples \footnote{For example the vector/one-form field of \cite{Bekenstein:2004ne} is necessarily Lorentz violating whereas the vector/one-form field of \cite{Moffat:2011rp} is massive.}.
The scalar and one-form perturbations in the present model may be stable in the sense of having right-sign mass-squared term and right-sign kinetic terms for a subregion of the $\{a_{1},b_{1},c_{1}\}$ parameter space. Collectively then the constraints on the parameter space are:

\begin{eqnarray}
\bar{u}^{2}\bp \left(b_{1}-  2\frac{a_{1}}{\bp^{2}}\right) &>& 0 \\
\bp \left(5b_{1}-2c_{1}\bp^{2}\right) &>& 0 \\
40a_{1}b_{1}\bp^{2} - 100a_{1}^{2}-3b_{1}^{2}\bp^{2} &>& 0
\end{eqnarray}
Recalling that $\bp = \bp(a_{1},b_{1},c_{1})$, we may use the background field equations to express $c_{1} = c_{1}(a_{1},b_{1},\bp)$ and express the above conditions as:

\begin{eqnarray}
 \bar{u}^{2}\bp \left(b_{1}\bp^{2}-  2a_{1}\right) &>& 0 \label{1stcon} \\
\bp \left(b_{1}\bp^{2}-10a_{1}\right) &<& 0 \label{2ndcon}\\
40a_{1}b_{1}\bp^{2} - 100a_{1}^{2}-3b_{1}^{2}\bp^{2} &>& 0 \label{3rdcon}
\end{eqnarray}
From the constraint (\ref{3rdcon}) we have that

\begin{eqnarray}
b_{1}\bp^{2} <10a_{1} < 3 b_{1}\bp^{2} 
\end{eqnarray}
The constraint (\ref{2ndcon}) then implies that $\bp >0$ whilst (\ref{1stcon}) provides no further constraint on the parameter space.

Though the above stability result appears encouraging, it should be noted that the full number of degrees of freedom that the theory possesses is as-yet unknown. It is conceivable that there exist additional degrees of freedom that are only excited around less simple background solutions or at higher order in perturbations. For example, note that - 
defining $\delta h^{IJ}$ via $\delta h^{I} = \delta h^{IJ}\bar{e}_{J}$ - only the antisymmetric part $\delta h^{[IJ]}$ appears in the equations of motion for linear perturbations around maximally symmetric backgrounds. If the theory contains additional degrees of freedom  then the results of this section may not be indicative of total perturbative stability. For example, one may calculate the Lagrangian of perturbations around these backgrounds to high enough order until a kinetic term for putative additional degrees of freedom appears; it may be that some of these kinetic terms are of the `wrong-sign', indicating a perturbative instability invisible to the analysis of this section.

 \section{Coupling to matter}
   \label{matter}
We now consider the coupling of matter fields when gravity is described by the pair $\{A^{AB},W^{AB}\}$. We will focus on the limit where the matter fields do not back-react on the gravitational field. This is the analogue of considering the formulation of matter fields for a fixed background frame $e^{I}$ in General Relativity. We will show that when gravity is regarded as a spontaneously-broken theory of $SU(2,2)\simeq SO(2,4)$ then all known matter fields can be cast in a first-order formalism without introducing auxiliary fields. It will be shown that the familiar Lagrangians for matter gauge fields and scalar fields which yield second-order field equations follow from being able to solve for some of the first-order variables in terms of others. We have already seen evidence of this possibility in the previous section wherein the perturbation $\delta \omega^{IJ}$ to the spin-connection could be solved for in terms of $\partial \delta e$ and $\partial \delta \alpha$. Similarly, the perturbation $\delta W^{IJ}$ was found to be fixed in terms of $\partial \delta c$.%

First we will show how familiar kinetic terms in the Lagrangian formalism are recovered for matter gauge fields, scalar Higgs fields, and spinor fields. Following this we will look at the type of potential terms that may be constructed for Higgs and spinor fields.
Throughout the section we will take the limit of the gravitational theory where 

\begin{align}
W^{ab}  &= \bp \epsilon^{ab}  \quad\quad (\bp = cst.) \\
W^{IJ} &= 0 \\
 DW^{Ia} &= e^{I}u^{a} \\
%\omega_{\mu}^{\ph{\mu}IJ} &=& 2 e^{\nu [I}\partial_{[\mu}e_{\nu]}^{J]}+ e_{\mu K} e^{\nu I} e^{\alpha J}\partial_{[\alpha}e_{\nu]}^{K} \\
\omega_{\mu}^{\ph{\mu}IJ} &= 2 e^{\nu [I}\partial_{[\mu}e_{\nu]}^{J]}+ e_{\mu K} e^{\nu I} e^{\alpha J}\partial_{[\alpha}e_{\nu]}^{K} \nn\\
& +  \frac{1}{u^{2}}e^{\nu [I}e_{[\nu}^{J]}\partial_{\mu]}u^{2}+ \frac{1}{2u^{2}}e_{\mu K} e^{\nu I} e^{\alpha J}e^{K}_{[\nu}\partial_{\alpha]}u^{2}  \\
c_{\mu} &=   -\frac{1}{u^{2}}u_{a}\epsilon^{a}_{\ph{a}b}\partial_{\mu}u^{b} \label{cond7}
\end{align}
where $\bar{\phi}$ is taken to be a constant and recall that $u^{2}\equiv \eta_{ab}u^{a}u^{b}$. This form for the fields $\{A^{AB},W^{AB}\}$ coincides with that taken for the maximally symmetric solutions 
of Section \ref{maxsym} but more generally we can assume that - much as for the Conformal Einstein-Cartan theory - this is a good approximation to the form the fields will take in any limit where the dynamics of the gravitational field is approximately that of General Relativity (as is undoubtedly the case on Earth). We assume that $u^{2}\neq 0$. The ansatz (\ref{cond7}) is invariant under local rescalings $u^{a}\rightarrow e^{-\alpha(x)}u^{a}$, $e^{I}\rightarrow e^{\alpha(x)}e^{I}$ and so we may set $u^{2} = \pm 1$ (depending on whether $u^{2}>0$ or $u^{2}<0$) if convenient to do so.
We will refer to this specific assumed form of the gravitational fields as the General-Relativistic limit of the full theory. 

\subsection{Adjoint Higgs, non-gravitational gauge fields}
\label{adjhiggs}
The result that part of a Higgs field (namely $\delta W^{IJ}$) enabled the recovery of a familiar Maxwell-like kinetic term for $\delta c$ is rather surprising and we will now show the same behaviour is repeated in the matter sector: it will be shown that the presence of a scalar field in the adjoint representation of a group ${\cal G}$ is necessary to yield familiar second-order dynamics of the gauge fields of that group.

Consider the $SO(2,4)$ model of gravity coupled to matter in the adjoint representation of some Lie group ${\cal G}$. We will take the field to be in the adjoint representation of ${\cal G}$ \emph{and} of the gravitational gauge group $SO(2,4)$ i.e. we consider a field with index structure

\begin{eqnarray}
\Phi^{A\ph{B}i}_{\ph{A}B\ph{i}j} \nn
\end{eqnarray}
where $i,j,k,\dots$ are ${\cal G}$ indices in the fundamental representation. This is an immediate departure from models such as the Einstein-Cartan theory wherein spacetime scalar fields have no `gravitational' indices. Henceforth for notational compactness we will suppress ${\cal G}$ indices and in the General-Relativistic limit we may express $\Phi^{AB}$ as follows:
 \begin{equation}
  \Phi^{AB} = \left( \begin{array}{cc}
\Phi^{IJ} & \frac{1}{u^{2}}Y^{I}u^{a}+\frac{1}{v^{2}}Z^{I}v^{a} \\
-\Phi^{Ia}  & \Phi \epsilon^{ab}
 \end{array} \right)
 \end{equation}
where $v^{a}\equiv \epsilon^{a}_{\ph{a}b}u^{b}$. Each of $\{\Phi, Y^{I},Z^{I},\Phi^{IJ}\}$ then transform in the adjoint representation of ${\cal G}$ but, respectively, as Lorentz scalar, Lorentz vectors, and anti-symmetric Lorentz tensor. Consider then the following locally $SO(2,4)\times {\cal G}$ invariant action, at most quadratic in $\Phi$ and ${\cal F} \equiv d{\cal B} + {\cal B}{\cal B}$, where ${\cal B}_{\mu}$ is the ${\cal G}$ group gauge field:

\begin{align}
S &=  - \int \epsilon_{ABCDEF}DW^{A}_{\ph{A}G}DW^{GB}\mathrm{Tr}\left(W^{CD}\Phi^{EF}{\cal F}+DW^{CD}\Phi^{E}_{\ph{E}H}D\Phi^{HF}\right) \label{1act}
\end{align}
where the trace denotes contraction of internal adjoint indices using the ${\cal G}$ Killing metric and  the $D$ operators are the full $SO(2,4)\times {\cal G}$ covariant derivatives. In the General-Relativistic limit, it will be useful to decompose the effect of $D$ on $\Phi^{AB}$ into the effect of the $SO(1,3)\times SO(1,1)\times {\cal G}$ covariant derivative ${\cal D}$ on $\Phi^{AB}$ plus any additional terms that may exist. It may be calculated that:

\begin{equation}
D\Phi^{AB} = \left( \begin{array}{cc}
 {\cal D}\Phi^{IJ} +\frac{2}{\bp}e^{[I}Z^{J]}  & {\cal D}\Phi^{Ia}  +\frac{1}{\bp} e^{I}u^{a}\Phi+ \frac{1}{\bp}v^{a}e_{J}\Phi^{IJ} \\
-D\Phi^{Ia} &{\cal D}\Phi^{ab} +\frac{2}{\bp u^{2}}v^{[a}u^{b]}e_{J}
Y^{J} 
 \end{array}\right)
\end{equation}
In this limit, the above action becomes (making the traces over  ${\cal G}$ indices implicit for now for notational compactness):

  \begin{align}
S &=   \int \left(\right. u^{2}Y^{M}{\cal D}_{M}\Phi
+\frac{u^{2}}{2\bp} Y^{M}Y_{M} -2\frac{u^4}{\bp}\Phi\Phi\nn\\
&+u^{2}\Phi^{MN}{\cal D}_{M}Z_{N}
-\frac{3}{2}\frac{u^{2}}{\bp}Z^{M}Z_{M}+\frac{1}{4\bp}\Phi^{MN}\Phi_{MN}+\frac{\bp u^{2}}{6}\Phi^{MN}{\cal F}_{MN}\left.\right)\epsilon_{IJKL}e^{I}e^{J}e^{K}e^{L}
 \end{align}
where we have used results from Appendix \ref{useful} to identify some terms as equal to one another. Varying with respect to $Y^{I}$ and $\Phi^{IJ}$ we obtain the equations of motion:
  \begin{align}
  0 &=   {\cal D}_{I}\Phi+ \frac{1}{\bp}Y_{I} \label{yeq} \\
 0 &=  u^{2}{\cal D}_{[I}Z_{J]}+\frac{1}{2\bp}\Phi_{IJ}+\frac{\bp u^{2}}{6}{\cal F}_{IJ} \label{pheq}
 \end{align}
 where ${\cal D}_{I} \equiv e^{\mu}_{I}{\cal D}_{\mu}$. The equations of motion (\ref{yeq}) and (\ref{pheq}) enable us to solve for $Y_{I}$ and $\Phi_{IJ}$ in terms of other fields; we may then use these expressions in the action, eliminating these fields from the action principle, yielding:
 
    \begin{align}
S &=   \int \left( -\frac{u^{2}\bp}{2}{\cal D}^{I}\Phi{\cal D}_{I}\Phi
 -\frac{2u^4}{\bp}\Phi\Phi \right.\nn\\
 &\left.-\frac{\bp^{3}}{36}\left({\cal F}^{IJ}+\frac{6}{\bp} {\cal D}^{[I}Z^{J]}\right)\left({\cal F}_{IJ}+\frac{6}{\bp} {\cal D}_{[I}Z_{J]}\right)
-\frac{3}{2}\frac{u^{2}}{\bp}Z^{I}Z_{I}\right)\epsilon_{IJKL}e^{I}e^{J}e^{K}e^{L} \label{sbz}
 \end{align}
We may now redefine the ${\cal G}$ gauge field ${\cal B}_{\mu}$, creating a new field:

\begin{eqnarray}
\tilde{\cal B}_{\mu} &\equiv &  {\cal B}_{\mu}  +  \frac{12}{\bp}Z_{\mu} \label{beeredef}
\end{eqnarray}
where $Z_{\mu} \equiv e_{\mu}^{I}Z_{I}$. Hence:

\begin{eqnarray}
{\cal F}_{\mu\nu} +\frac{6}{\bp}{\cal D}_{[\mu}Z_{\nu]}  =  \tilde{\cal F}_{\mu\nu}-\frac{288}{\bp^{2}}Z_{[\mu}Z_{\nu]}
\end{eqnarray}
Then, varying with respect to $Z_{I}$ we obtain an equation polynomial in $Z_{I}$. Contributions to this equation that cause there to be only non-zero solutions for $Z_{I}$ arise from the coupling of Higgs and fermionic fields to $Z_{I}$ that emerge from the gauge field redefinition (\ref{beeredef}). Inserting this solution back into the action (\ref{sbz}) then yields the following

\begin{eqnarray}
S &=&   \int \mathrm{Tr}\left( -\frac{\bp^{3}}{36}\tilde{\cal F}^{MN}\tilde{\cal F}_{MN}-\frac{u^{2}\bp}{2}{\cal D}^{M}\Phi{\cal D}_{M}\Phi
 -\frac{2u^4}{\bp}\Phi\Phi + \dots
\right)\epsilon_{IJKL}e^{I}e^{J}e^{K}e^{L} \label{act7}
 \end{eqnarray}
where ellipses denote terms higher order in Higgs, fermion (expected to appear via their original coupling to ${\cal B}_{\mu}$), and gauge fields.
We can easily cast this into the metric formalism, defining $g_{\mu\nu} \equiv \eta_{IJ}e^{I}_{\mu}e^{J}_{\nu}$ we may write (\ref{act7}) as:

\begin{eqnarray}
S &=&   -24\int \mathrm{Tr}\left( \frac{\bp^{3}}{36}\tilde{\cal F}^{\mu\nu}\tilde{\cal F}_{\mu\nu}+\frac{u^{2}\bp}{2}{\cal D}^{\mu}\Phi{\cal D}_{\mu}\Phi
 +\frac{2u^4}{\bp}\Phi\Phi + \dots
\right)\sqrt{-g}d^{4}x \label{act8}
 \end{eqnarray}
Therefore, second-order kinetic terms for a gauge field $\tilde{B}$ and massive adjoint Higgs field $\Phi$ are recovered from the first-order, polynomial action (\ref{1act}) built from the pair $\{{\cal B},\Phi^{AB}\}$. The mass term for $\Phi$ arises from the coupling of $\Phi^{AB}$ to the full $SU(2,2)\times {\cal G}$-covariant derivative $D$. This is rather like the origin of mass terms for gauge bosons in gauge theory coming via covariant derivatives of Higgs fields, but here because the `gauge boson' is the spacetime frame $e^{I}$ itself, the more direct interpretation is that such a term provides a mass for the Higgs field itself. Interestingly, the kinetic terms for $\tilde{\cal B}$ and $\Phi$ are of the correct relative sign only if $u^{2}>0$.

\subsection{Fundamental representation Higgs}
\label{funhiggs}
We now describe the coupling of gravity to a Higgs field valued in the fundamental representation of ${\cal G}$. We will look to construct a first-order formalism for a field $\varphi^{A}$ which is in the fundamental representation of ${\cal G}$ \emph{and} $SO(2,4)$ i.e.

\begin{eqnarray}
\varphi^{Ai} \nn
\end{eqnarray}
where we use $i,j,k,\dots$ for indices in the fundamental representation of ${\cal G}$. Again this contrasts with the coupling of gravity to fundamental representation Higgs fields in the Einstein-Cartan theory wherein the scalar field has no gravitational indices. For concreteness we will look at the case where ${\cal G} = SU(N)$. This enables us to construct the field $\phi^{\dagger A}_{i} \equiv (\phi^{Ai'})^{*}\delta_{i'i}$ where $\delta_{i'i}$ is the $SU(N)$ invariant matrix. Generalization to other groups such as $SO(N)$ is straightforward, instead requiring use of the $SO(N)$-invariant matrix in the kinetic term. Consider the following action:

\begin{eqnarray}
S = \int  \epsilon_{ABCDEF}DW^{A}_{\ph{A}G}DW^{GB}DW^{CD}\varphi^{\dagger E}D\varphi^{F}\label{awgravity}
\end{eqnarray}
Again it is useful to decompose the full $SU(2,2)\times {\cal G}$-covariant derivative into the 
$SO(1,3)\times SO(1,1)\times {\cal G}$ covariant derivative ${\cal D}$ and additional terms:

\begin{eqnarray}
  D\varphi^{I}   &=&  {\cal D}\varphi^{I} -\frac{1}{\bp} e^{I}v_{a}\varphi^{a}\\
D\varphi^{a} &=& {\cal D}\varphi^{a} +\frac{1}{\bp}v^{a}e_{I} \varphi^{I}
\end{eqnarray}
As before, we restrict ourselves to the General-Relativistic limit, and decompose $\varphi^{A}$
as follows:

 \[ \varphi^{A} = \left( \begin{array}{cc}
\varphi^{I} \\
\frac{1}{u^{2}}\varpi u^{a}+\frac{1}{v^{2}}\varphi v^{a}
 \end{array} \right)\]
and the action may be shown to take the form:

\begin{align}
S &= \int -2u^{2}\epsilon_{IJKL}e^{I}e^{J}e^{K}(\varphi^{L}\bar{\cal D}\varphi^{\dagger} + \varphi^{\dagger L}\bar{\cal D}\varphi) +\frac{2u^{4}}{\bp}\epsilon_{IJKL}e^{I}e^{J}e^{K}\varphi^{\dagger L}e_{M}\varphi^{M} \nn \\
&-2u^{2}\frac{\varphi^{\dagger}\varphi}{\bp}\epsilon_{IJKL}e^{I}e^{J}e^{K}e^{L}\nn \\
%&=& \int -2u^{2}\epsilon_{IJKL}g^{I}g^{J}g^{K}(\varphi^{L}g_{M}({\cal D}^{M}\varphi^{*}-\frac{u^{2}}{2}\varphi^{*M}) + \varphi^{*L}g_{M}({\cal D}^{M}\varphi
%-\frac{u^{2}}{2}\varphi^{M}))  \\
&= \int -4u^{2}\epsilon_{IJKL}e^{I}e^{J}e^{K}\varphi^{\dagger L}e_{M}(\bar{\cal D}^{M}\varphi
-\frac{u^{2}}{2\bp}\varphi^{M})  -2u^{2}\frac{\varphi^{\dagger}\varphi}{\bp}\epsilon_{IJKL}e^{I}e^{J}e^{K}e^{L}
\end{align}
Where 

\begin{align}
\tilde{\cal D}\phi \equiv {\cal D}\phi - \frac{\phi}{2u^{2}}du^{2}
\end{align}
Note that in the General-Relativistic limit, the field $\varpi$ disappears from the action. Varying with respect to $\phi^{\dagger L}$ we recover:

\begin{eqnarray}
0 &=&  \epsilon_{IJKL}e^{I}e^{J}e^{K}e_{M}(\tilde{\cal D}^{M}\varphi-\frac{u^{2}}{2\bp}\varphi^{M})
   - \frac{u^{2}}{2\bp} \epsilon_{IJKM}e^{I}e^{J}e^{K}\varphi^{M}e_{L}
\end{eqnarray}
We can solve this equation to yield: $\varphi^{M} =\frac{\bp}{u^{2}}\tilde{\cal D}^{M}\varphi $. Inserting this back into the action yields:

\begin{align}
S &= \int -\frac{\bp}{2}\left( \tilde{\cal D}^{M}\varphi^{\dagger} \tilde{\cal D}_{M}\varphi\right) \epsilon_{IJKL}e^{I}e^{J}e^{K}e^{L}-2u^{2}\frac{\varphi^{\dagger}\varphi}{\bp}\epsilon_{IJKL}e^{I}e^{J}e^{K}e^{L}\nn\\
 &= -24\int\left( \frac{\bp}{2}\left( \tilde{\cal D}^{\mu}\varphi^{\dagger} \tilde{\cal D}_{\mu}\varphi\right) +2\frac{u^{2}}{\bp}\varphi^{\dagger}\varphi\right)\sqrt{-g}d^{4}x \label{funhig}
\end{align}
Therefore in the General-Relativistic limit we recover the action for a massive scalar field. The mass-squared of the scalar field is real if $u^{2}>0$. Though we have focused on the case where the scalar field is in the fundamental representation of ${\cal G}$ (and of $SO(2,4)$), we do not see an obstruction to recovering second-order dynamics from similar actions to (\ref{awgravity}) for Higgs fields that may exist in other representations of ${\cal G}$ as long as ${\cal G}$ possesses structure so that $SU(2,2)\times {\cal G}$ invariant actions can be constructed. 

\subsection{Spinor fields}
\label{spinors}
So far we have dealt exclusively with real representations of the group $SO(2,4)$. To incorporate spinorial matter into the theory it is necessary to make use of the complex representations of $SU(2,2)$. To recap, the group $SU(2,2)$ has a matrix representation as the set of $4\times 4$ matrices $U^{\alpha}_{\ph{\alpha}\beta}$ with unit determinant that preserve the Hermitian matrix 

\begin{eqnarray}
h_{\alpha'\alpha} = \left( \begin{array}{cc}
0 & I \\
I  & 0
 \end{array} \right)
\end{eqnarray}
where $I$ is the $2\times 2$ unit matrix. We can then define a four-dimensional complex vector space $\mathds{C}^{(2,2)}$ (we use the notation $\chi^{\alpha}$ to denote a vector in this space) and primed indices denote vectors belonging to the conjugate space $\mathds{C}^{*(2,2)}$ e.g. if a vector $\chi^{\alpha}$ transforms as $U^{\alpha}_{\ph{\alpha}\beta}\chi^{\beta}$ then a vector $\beta^{\alpha'}$ transforms as $(U^{*})^{\alpha'}_{\ph{\alpha'}\beta'}\beta^{\beta'}$. The space $\mathds{C}^{(2,2)}$ possesses the symmetric inner product $\left(,\right)$:

\begin{eqnarray}
(\beta,\chi) \equiv  \frac{1}{2}h_{\alpha'\alpha}\left((\beta^{*})^{\alpha'}\chi^{\alpha}+(\chi^{*})^{\alpha'}\beta^{\alpha}\right)
\end{eqnarray}
We may additionally consider spaces of `forms'. For example consider $\bigwedge^{2} \mathds{C}^{(2,2)}$, the space of antisymmetric matrices $\xi^{\alpha\beta} = -\xi^{\beta\alpha}$. 
This is a six-dimensional complex vector space equipped with symmetric inner product $\left<,\right>$:

\begin{equation}
\left< \mu, \xi \right >  \equiv \frac{1}{2}h_{\alpha'\alpha}h_{\beta'\beta}\left((\mu^{*})^{\beta'\alpha'}\xi^{\alpha\beta}+(\xi^{*})^{\beta'\alpha'}\mu^{\alpha\beta}\right)
\end{equation}
By definition the completely antisymmetric symbol $\epsilon_{\alpha\beta\gamma\delta}$ is invariant under $SU(2,2)$ transformations. The presence of this invariant symbol allows us to decompose elements according how they transform under operation by the antisymmetric symbol: a `real' matrix $\tilde{\xi}^{\alpha\beta}$ is one that satisfies:

\begin{eqnarray}
\tilde{\xi}_{\alpha\beta} =\frac{1}{2} \epsilon_{\alpha\beta\gamma\delta}(\tilde{\xi}^{*})^{\gamma\delta} \label{selfdual}
\end{eqnarray}
where indices on $(\xi^{*})^{\alpha'\beta'}$ have been raised with $h^{\alpha\alpha'}$. The space of matrices satisfying (\ref{selfdual}) thus has six real dimensions and we may express a given $\xi^{\alpha\beta}$ in terms of a basis $\sigma_{A}^{\ph{A}\alpha\beta}$ :

\begin{eqnarray}
\tilde{\xi}^{\alpha\beta}  = \tilde{\xi}^{A} \sigma_{A}^{\ph{A}\alpha\beta}
\end{eqnarray}
where the coefficients $\tilde{\xi}_{A}$ are real numbers and:

\begin{eqnarray}
\sigma_{A\alpha\beta} =\frac{1}{2} \epsilon_{\alpha\beta\gamma\delta}\sigma^{*\gamma\delta}_{A}
\end{eqnarray}
An explicit set of $\sigma_{A\alpha\beta}$ are:

\begin{eqnarray}
\sigma_{(-1)} &=&  \left( \begin{array}{cccc}
0 & 1 & 0 & 0\\
-1  & 0 & 0 & 0\\
0 & 0 & 0 & 1 \\
0 & 0 & -1 & 0 \\
 \end{array} \right)  , \quad
  \sigma_{(0)} = \left( \begin{array}{cccc}
0 & 0 & 0 & i\\
0  & 0 & -i & 0\\
0 & i & 0 & 0 \\
-i & 0 & 0 & 0 \\
 \end{array} \right) \nn\\
  \sigma_{(1)} &=& 
 \left( \begin{array}{cccc}
0 & 0 & -i & 0\\
0  & 0 & 0 & i\\
i & 0 & 0 & 0 \\
0 & -i & 0 & 0 \\
 \end{array} \right), \quad
  \sigma_{(2)} =  \left( \begin{array}{cccc}
0 & 0 & 1 & 0\\
0  & 0 & 0 & 1\\
-1 & 0 & 0 & 0 \\
0 & -1 & 0 & 0 \\
 \end{array} \right) \nn \\
  \sigma_{(3)} &=&  \left( \begin{array}{cccc}
0 & 0 & 0 & i\\
0  & 0 & i & 0\\
0 & -i & 0 & 0 \\
-i & 0 & 0& 0 \\
 \end{array} \right) 
 , \quad
 \sigma_{(4)} = \left( \begin{array}{cccc}
0 & -1 & 0 & 0\\
1  & 0 & 0 & 0\\
0 & 0 & 0 & 1 \\
0 & 0 & -1 & 0 \\
 \end{array} \right) 
 \end{eqnarray}
It can be checked then that 

\begin{eqnarray}
\frac{1}{2}\left( \sigma_{A}^{*\alpha\gamma}\sigma_{B\gamma\beta}+\sigma_{B}^{*\alpha\gamma}\sigma_{A\gamma\beta}\right) = \eta_{AB} \delta^{\alpha}_{\ph{\alpha}\beta} \label{metr}
\end{eqnarray}
where $\eta_{AB}=\mathrm{diag}(-1,-1,1,1,1,1)$ is the invariant matrix  of $SO(2,4)$. Indeed, we have that:

\begin{equation}
\left< \tilde{\mu}, \tilde{\xi} \right >  =\frac{1}{2} \tilde{\mu}^{A}\tilde{\xi}^{B} h_{\alpha'\alpha}h_{\beta'\beta}\left(\sigma^{*\beta'\alpha'}_{A}\sigma_{B}^{\alpha\beta} +\sigma^{*\beta'\alpha'}_{B}\sigma_{A}^{\alpha\beta}\right) = 4 \eta_{AB}\tilde{\mu}^{A}\tilde{\xi}^{B} 
\end{equation}
We can write an $SU(2,2)$ group element $U^{\alpha}_{\ph{\alpha}\beta}$ as:

\begin{eqnarray}
U^{\alpha}_{\ph{\alpha}\beta} =  \left(e^{i \theta_{j}T^{j}} \right)^{\alpha}_{\ph{\alpha}\beta}
\end{eqnarray}
The unitarity of $U$ implies that the generators $(T^{j})^{\alpha}_{\ph{\alpha}\beta}$ must then satisfy $(T^{j})_{\alpha'\alpha} =  (T^{j \dagger})_{\alpha'\alpha}$ i.e. they are Hermitian when indices have been lowered with $h_{\alpha'\alpha}$. Consider the set of fifteen matrices:

\begin{eqnarray}
(j_{AB})^{\alpha}_{\ph{\alpha}\beta} &=&\frac{i}{4}\left( \sigma_{A}^{*\alpha\gamma}\sigma_{B\gamma\beta}-\sigma_{B}^{*\alpha\gamma}\sigma_{A\gamma\beta}\right)\\
&=& \frac{i}{2}\left(\sigma_{A}^{*\alpha\gamma}\sigma_{B\gamma\beta}-\eta_{AB}\delta^{\alpha}_{\ph{\alpha}\beta}\right)
\end{eqnarray}
where we have used the result (\ref{metr}). It can be checked that the matrices $j_{AB}$ satisfy the following Lie algebra:

\begin{eqnarray}
[j^{AB},j^{CD}] &=&  i\left(\eta^{BC}j^{AD}-\eta^{AC}j^{BD}-\eta^{BD}j^{AC}+\eta^{AD}j^{BC}\right)
\end{eqnarray}
This indeed is the Lie algebra of $SU(2,2)$ and $SO(2,4)$. To put things on a more familiar footing, the generators $j_{AB}$ are explicitly given by:
\begin{eqnarray}
j^{-1,I} &=& \frac{1}{2} \gamma^{I} , \quad  j^{4,I} = \frac{1}{2}\gamma_{5}\gamma^{I} \\
j^{IJ} &=& \frac{1}{4i}[\gamma^{I},\gamma^{J}] , \quad j^{-1,4} = i\gamma^{5}
\end{eqnarray}
where

 \begin{equation} 
 \gamma^{I} = \left( \begin{array}{cc}
0 & \Sigma^{I} \\
\bar{\Sigma}^{I} & 0
 \end{array} \right)     ,\quad  \gamma_{5}  = \left( \begin{array}{cc}
-1 & 0 \\
0& 1
 \end{array} \right)
 \end{equation}
where $\Sigma^{I} = (1, \Sigma^{i})$, $\bar{\Sigma}^{I} = (1,-\Sigma^{i})$, where $\Sigma^{i}$ are the Pauli matrices.
We see then that $j^{IJ}$ generate Lorentz transformations and under these transformations an $SU(2,2)$ spinor $\chi^{\alpha}$ transforms like a four-component
representation of $Spin(1,3)$ i.e. a \emph{Dirac} spinor. We can additionally relate an $SU(2,2)$ ${\cal A}^{\alpha}_{\ph{\alpha}\beta}$ connection to the $SO(2,4)$ connection $A^{A}_{\ph{A}B}$ used so far:

\begin{eqnarray}
{\cal A}^{\alpha}_{\ph{\alpha}\beta} &=&  \frac{1}{2} A_{AB}j^{AB\alpha}_{\ph{AB\alpha}\beta}
\end{eqnarray}
The $SU(2,2)$-covariant derivative of $\chi^{\alpha}$ is then as follows:

\begin{eqnarray}
D\chi^{\alpha}  &=& d\chi^{\alpha}  -i{\cal A}^{\alpha}_{\ph{\alpha}\beta} \chi^{\beta} \\
 &=& d\chi^{\alpha}  -\frac{i}{2} A_{AB}j^{AB\alpha}_{\ph{AB\alpha}\beta}
 \chi^{\beta} 
\end{eqnarray}
If $\chi$ additionally belongs to a representation of a group ${\cal G}$ then we can decompose the derivative $D$ into the $SO(1,3)\times SO(1,1)\times {\cal G}$ covariant derivative ${\cal G}$ and additional pieces:

\begin{eqnarray}
D\chi^{\alpha} &=& {\cal D}\chi^{\alpha}  -\frac{i}{2} A_{-1J}j^{-1J\alpha}_{\ph{-1J\alpha}\beta}\chi^{\beta}-\frac{i}{2} A_{4J}j^{4J\alpha}_{\ph{4J\alpha}\beta}\chi^{\beta}\\
&=& {\cal D}\chi^{\alpha}  -\frac{i}{2} A_{-1J}\gamma^{J\alpha}_{\ph{J\alpha}\beta}\chi^{\beta}-\frac{i}{2} A_{4J}(\gamma^{5}\gamma^{J})^{\alpha}_{\ph{\alpha}\beta}\chi^{\beta}
\end{eqnarray}
If we take the case $u^2>0$ and fix $u^{2} = 1$ then we may choose a gauge where $u^{a} \overset{*}{=} \delta^{a}_{4}$, hence $v^{a} \overset{*}{=}-\delta^{a}_{-1}$ and 
$A^{I,-1} \overset{*}{=} \frac{1}{\bp}e^{I}$, $A^{I4} \overset{*}{=}0$ hence:

\begin{eqnarray}
D\chi^{\alpha} &\overset{*}{=}& {\cal D}\chi^{\alpha}  -\frac{i}{2\bp}e_{J}\gamma^{J\alpha}_{\ph{J\alpha}\beta}\chi^{\beta}
\end{eqnarray}
We can now write down an action for a spinor field coupled to gravity:

\begin{eqnarray}
S =  i\int \epsilon_{ABCDEF}DW^{A}_{\ph{A}G}DW^{GB}DW^{CD}\bar{\chi}j^{EF}D\chi \label{spinact1}
\end{eqnarray}
where $\bar{\chi}_{\alpha} \equiv (\chi^{*})^{\alpha'}h_{\alpha'\alpha}$ and we implicitly use any invariant group structure from ${\cal G}$ necessary for (\ref{spinact1}) to be ${\cal G}$-invariant. In the General-Relativistic limit this action takes the form:

\begin{eqnarray}
S \overset{*}{=} - i\int 4 \epsilon_{IJKL}e^{I}e^{J}e^{K}\bar{\chi}j^{La}v_{a}D\chi
\end{eqnarray}
We have that $v_{a} \overset{*}{=} \delta^{-1}_{a}$ and so $j^{La}v_{a} \overset{*}{=} j^{L,-1}\overset{*}{=} -\frac{1}{2}\gamma^{L}$, hence:

\begin{align}
S &\overset{*}{=}  i\int 2 \epsilon_{IJKL}e^{I}e^{J}e^{K}\bar{\chi}\gamma^{L}D\chi =  i\int 2\epsilon_{IJKL}e^{I}e^{J}e^{K}\bar{\chi}\gamma^{L}({\cal D}\chi -\frac{i}{2\bp}e_{M}\gamma^{M}\chi)\nn \\ 
&=   \int 2i \epsilon_{IJKL}e^{I}e^{J}e^{K}\bar{\chi}\gamma^{L}{\cal D}\chi -\frac{1}{\bp}\bar{\chi}\chi\epsilon_{IJKL}e^{I}e^{J}e^{K}e^{L}  \label{szpinak}
\end{align} 
Note that in this $SO(1,1)$ gauge we can see from equation (\ref{cond7}) that  $c\overset{*}{=}0$. Thus we see that in the General-Relativistic limit, the action (\ref{spinact1}) reduces to the action for a massive Dirac spinor covariant under local $Spin(1,3)\times {\cal G}$ transformations.

\subsection{Chirality and mass}
\label{chirmass}
We have seen that it is possible to recover second-order dynamics for gauge and Higgs fields from a first-order perspective. Additionally, we see that mass terms for Higgs fields $\Phi$, $\varphi$, and spinor fields $\chi$ appear quite naturally in the context of first-order $SU(2,2)\times {\cal G}$-invariant actions via the $SU(2,2)$ covariant derivative.

As spinorial representations of $SU(2,2)$, fermionic fields are necessarily four-component Dirac spinor fields. Compare this to the standard model of particle physics wherein fermions are two-component Weyl spinors. Though prior to the standard model it was thought that the left-handed electron $E_{L}$ and right-handed electron $E_{R}$ were indeed two parts of a single Dirac spinor $\Psi = (E_{L},E_{R})$, it is now known that this is not the correct structure. Rather, $E_{R}$ is a Weyl spinor and an $SU(2)$ \emph{singlet} whilst $E_{L}$ is part of an $SU(2)$ \emph{doublet} Weyl spinor (the electron-neutrino), and so $E_{L}$ couples to $SU(2)$ gauge fields directly via the covariant derivative, whereas $E_{R}$ does not. A chirality transformation is taken to interchange $E_{R} \leftrightarrow E_{L}$. Clearly the standard model Lagrangian cannot be invariant under this transformation. In this sense the standard model is referred to as being a chiral theory.

Focusing on the $SU(2)$ singlet $E_{R}$, from the $SU(2,2)$ perspective there are no two-dimensional spinorial representations 
this field must be part of a Dirac spinor $\chi = ({\cal M}_{L},E_{R})$ where ${\cal M}_{L}$ has the same hypercharge as $E_{R}$ but transforms as a left-handed Weyl spinor under Lorentz transformations (these defined post-symmetry breaking by $W^{AB}$). The field ${\cal M}_{L}$ cannot be identified with the left-handed electron because it possesses the wrong $SU(2)$ index structure. Indeed, the left-handed electron-neutrino doublet  will be part of another $SU(2,2)$ spinor additionally containing a right-handed $SU(2)$ doublet with the same $SU(2)\times U(1)$ hypercharges but transforming as a right-handed Weyl spinor. Where are these new fields in nature? 

If the action for the dynamics of $\chi =({\cal M}_{L},E_{R})$ were described by (\ref{szpinak}) then we have a symmetry under interchange of ${\cal M}_{L}\leftrightarrow E_{R}$; thus ${\cal M}_{L}$ would have the same mass as $E_{R}$. 
The field ${\cal M}_{L}$ has not been observed in nature and so if the $SU(2,2)$ approach is a viable model of gravitation there must be an explanation for this.
If it is the case that ${\cal M}_{L}$ is simply too massive to have been detected yet, then there must exist additional terms in the spinor action that break the ${\cal M}_{L}\leftrightarrow E_{R}$ symmetry. 
Consider the following action involving $\chi = ({\cal M}_{L},E_{R})$:

\begin{eqnarray}
S &=&   \int  \epsilon_{ABCDEF} \bar{\chi} j^{AB}\chi DW^{C}_{\ph{C}G}DW^{GD}DW^{E}_{\ph{E}H}DW^{HF} \nn \\
 &= &  -2i\int  \bar{\chi}\gamma_{5}\chi \epsilon_{IJKL} e^{I}e^{J}e^{K}e^{L} \label{grzyb}
\end{eqnarray}
where the second equality applies only in the General-Relativistic limit and we have fixed $u^{2}=1$.
Under ${\cal M}_{L}\leftrightarrow E_{R}$, the action (\ref{grzyb}) becomes the minus of its original value whereas the action (\ref{szpinak}) is unchanged. If $\chi$ were described by the combined actions (\ref{szpinak}) and (\ref{grzyb}) then its action will not transform
homogeneously under ${\cal M}_{L}\leftrightarrow E_{R}$ and generally ${\cal M}_{L}$ and $E_{R}$ will have different masses.
Of course this is some way from showing that it is to be expected that `mirror' fields like ${\cal M}_{L}$ to be unobserved whilst retaining the physics of the standard model at lower energies, but clearly there is structure in the gravitational sector that will typically prevent the familiar field $E_{R}$ and unfamiliar field ${\cal M}_{L}$ having the same mass.

Another source of fermion mass should be via Yukawa-type interactions with a Higgs field such as $\varphi^{A}$. Consider a gauge group ${\cal G}$, assumed to have a matrix representation. Using $i,j,k,\dots $ for indices in the fundamental representation of ${\cal G}$ (e.g. the fundamental representation Higgs of $SU(5)$ would have the $SU(2,2)\times {\cal G}$ index structure $\varphi^{Ai}$) then we may write down an action coupling two separate $SU(2,2)$ spinors $P^{i \alpha}$  and $E^{\alpha}$. Consider the following action:

\begin{eqnarray} 
S &=& \int \epsilon_{ABCDEF} \varphi^{*E}_{i} P^{i\alpha}\sigma^{F}_{\alpha\beta} E^{\beta}  DW^{A}_{\ph{C}G}DW^{GB}DW^{C}_{\ph{E}H}DW^{HD}\nn\\
&= & \int \epsilon_{ab}v^{a}\varphi^{*}_{i} P^{i\alpha}\sigma^{b}_{\alpha\beta}E^{\beta} \epsilon_{IJKL}e^{I}e^{J}e^{K}e^{L} \nn\\
&\overset{*}{=}& - \int \varphi^{*}_{i} P^{i\alpha}C_{\alpha\beta}E^{\beta} \epsilon_{IJKL}e^{I}e^{J}e^{K}e^{L} \nn\\
&\equiv &  -  \int \varphi^{*}_{i}L^{\dagger i\beta'}h_{\beta'\beta}E^{\beta} \epsilon_{IJKL}e^{I}e^{J}e^{K}e^{L} \label{yak}
\end{eqnarray}
where the second equality applies only in the General-Relativistic limit, $C_{\alpha\beta}$ is the $Spin(1,3)$-invariant charge conjugation matrix recovered from projecting $\sigma^{a}_{\alpha\beta}$ along $u_{a}$, and we have defined the spinor $L^{i\beta}$ via

\begin{eqnarray}
L^{i'\beta} = P^{\dagger i'\alpha'}C_{\alpha'\beta'}h^{\beta'\beta}
\end{eqnarray}
For example, focusing specifically on the case where ${\cal G}$ is the electroweak group $SU(2)\times U(1)$, we see this is a Yukawa-type mass term for a left-handed field $SU(2)$-doublet  within $L^{i'\beta}$ coupled to a right-handed $SU(2)$ singlet within $E^{\beta}$ along with an identical term for a right-handed field $SU(2)$-doublet within $L^{i'\beta}$ coupled to a left-handed $SU(2)$ singlet within $E^{\beta}$. 

The term (\ref{yak}) acts as a mass term only when the field $\varphi$ achieves a non-vanishing expectation value. Suppressing internal indices for compactness again, we now briefly consider how potential terms may be constructed for fields such as $\Phi^{AB}$, $\varphi^{A}$ such that $\Phi$ and $\varphi$ achieve non-vanishing expectation values, thus spontaneously breaking the symmetry ${\cal G}$. 

In the General-Relativistic limit, and taking $u^{2}=1$, scalars formed with the field $W^{AB}$ reduce to familiar terms e.g.

\begin{eqnarray}
\mathrm{Tr}(W_{AB}W_{CD}\Phi^{AB}\Phi^{CD}) &=&  4\bp^{2}\mathrm{Tr}(\Phi\Phi)\\
W_{AB}W^{B}_{\ph{B}C}\varphi^{\dagger A}\varphi^{C} &=&  \bp^{2}\varphi^{\dagger}\varphi
\end{eqnarray}
Hence, polynomial functions of these scalars appearing alongside terms which are proportional to the familiar spacetime volume four-form in the General-Relativistic limit (e.g. the $c_{1}$ term from the gravitational action) will enable the recovery of symmetry breaking potentials.

Therefore we see that symmetry-breaking potentials for Higgs fields and masses for Weyl fermions and their `mirrors'  may be recovered from an $SU(2,2)$ framework in the General-Relativistic limit. 
\subsection{Field redefinitions and collected results}
\label{fieldredefs}
Collectively, the actions (\ref{act8}),(\ref{funhig}), and (\ref{szpinak}) for gauge field, Higgs fields, and spinor field then take the form):

\begin{align}
S &=   -24\int \mathrm{Tr}\left(\frac{\bp^{3}}{36}\tilde{\cal F}^{\mu\nu}\tilde{\cal F}_{\mu\nu}+\frac{u^{2}\bp}{2}{\cal D}^{\mu}\Phi{\cal D}_{\mu}\Phi
+\frac{2u^4}{\bp}\Phi\Phi  \right.  \nn \\
&\left.\right.+\frac{\bp}{2} \tilde{\cal D}^{\mu}\varphi^{\dagger} \tilde{\cal D}_{\mu}\varphi+2\frac{u^{2}}{\bar{\phi}}\varphi^{\dagger}\varphi -\left.\frac{i |u|u^{2}}{2} e^{\mu}_{L}\bar{\chi}\gamma^{L}{\cal D}_{\mu}\chi +\frac{u^{4}}{\bp}\bar{\chi}\chi+ \dots\right)\sqrt{-g}d^{4}x
\end{align}
where the dots denote terms higher order in fields. Here we have 
not fixed $|u^{2}|$ to be of unit magnitude but we restrict ourselves to the case $u^{2}>0$. Then, making the field redefinitions $\tilde{\Phi} = 3|u|\Phi/\bp$, $\tilde{\varphi}= 3\varphi/\bp$, $\tilde{\chi}=3\chi |u|^{3/2}/(\sqrt{2}\sqrt{\phi^{3}})$ and introducing the following derivatives

\begin{align}
\tilde{\cal D}\tilde{\Phi} &= {\cal D}\tilde{\Phi}- \frac{\tilde{\Phi}}{2u^{2}}du^{2} \\
\tilde{\cal D}\tilde{\chi} &= {\cal D}\tilde{\chi}- \frac{3\tilde{\chi} }{4u^{2}}du^{2}
\end{align}
we have in total that the action

\begin{align}
- &\int \epsilon_{ABCDEF}DW^{A}_{\ph{A}G}DW^{GB}\left(W^{CD}\mathrm{Tr}\left(\Phi^{EF}{\cal F}\right)\right.\nn\\
&\left.+DW^{CD}\left(\mathrm{Tr}\left(\Phi^{E}_{\ph{E}H}D\Phi^{HF}\right)-\varphi^{\dagger E}D\varphi^{F}-  i\bar{\chi} j^{EF}D\chi\right)\right) \label{wmatter}
\end{align}
yields field equations in the General-Relativistic limit equivalent to those obtained from the action 

\begin{align}
 \frac{8\bp^{3}}{3}&\int \left( -\frac{1}{4}\mathrm{Tr}(\tilde{\cal F}^{\mu\nu}\tilde{\cal F}_{\mu\nu})-\frac{1}{2}\mathrm{Tr}(\tilde{\cal D}^{\mu}\tilde{\Phi}\tilde{\cal D}_{\mu}\tilde{\Phi})
-\frac{1}{2} \tilde{\cal D}^{\mu}\tilde{\varphi}^{\dagger} \tilde{\cal D}_{\mu}\tilde{\varphi} +i e^{\mu}_{L}\bar{\tilde{\chi}}\gamma^{L}\tilde{\cal D}_{\mu}\tilde{\chi}\right. \nn \\
&\left.-\frac{m_{g}^{2}}{2}\mathrm{Tr}(\tilde{\Phi}\tilde{\Phi}) -\frac{m_{g}^{2}}{2}\tilde{\varphi}^{\dagger}\tilde{\varphi}  -m_{g}\bar{\tilde{\chi}}\tilde{\chi}+ \dots\right)\sqrt{-g}d^{4}x \label{actions}
\end{align}
where $m_{g}\equiv 2|u|/\bp$. Furthermore, the action (\ref{actions}) is invariant under the following local scale transformations:

\begin{align*}
e^{I}_{\mu} &\rightarrow e^{-\alpha(x)}e^{I}_{\mu}\\
u^{a} \rightarrow e^{\alpha(x)} u^{a} \quad   \tilde{\varphi} &\rightarrow e^{\alpha(x)} \tilde{\varphi} \quad 
\tilde{\Phi} \rightarrow e^{\alpha(x)}  \tilde{\Phi}\\
\tilde{\chi} &\rightarrow e^{\frac{3}{2}\alpha(x)}  \tilde{\chi}\\
\tilde{B}_{\mu} &\rightarrow \tilde{B}_{\mu} \\
\phi &\rightarrow \phi
\end{align*}
Thus we see that locally scale-invariant versions of standard matter kinetic terms are recovered alongside `gravitational' mass terms, each providing a contribution $m_{g}$ to field mass. The invariance of $\phi$ under the local scale transformation suggests that it should be treated as a dimensionless field. An informed estimate of $m_{g}$ would involve a simultaneous treatment of the gravitational behaviour of the theory, in particular how $\bp$ and $u^2$ are related to observed scales.

\section{Observable consequences}
\label{observable}
It is clear from the preceding sections that the $SU(2,2)$ approach to gravity will include degrees of freedom beyond those present when General Relativity is coupled to known matter fields. As follows we sketch how the model may have a role to play in accounting for various cosmological phenomena and how this could be tested.

\subsection{Inflaton Candidate}
In the standard model of cosmology early-time, large-scale perturbations to matter and part of the gravitational metric are taken to be of a specific form: they are (at least approximately) adiabatic \cite{Christopherson:2008ry} and near scale-invariant \cite{Mukhanov:1990me}. A popular explanation for this is that these  fields inherited these properties via coupling to a scalar field,  the inflaton, whose background evolution accelerates the early universe rapidly for a time and fluctuations of the field against this background lead to the above properties on large scales. Could the field $\phi$ be the inflaton?
We note that if the `alignment of frames' condition (\ref{80sstallone}) holds then, neglecting terms in the one-form $c$ we have the following contributions to the action (\ref{act5}):

\begin{align}
S[\phi,e^{I},\omega^{IJ}] &=  \frac{1}{32\pi G}\int\left( \epsilon_{IJKL}\tilde{e}^{I}\tilde{e}^{J}R^{KL}(\omega) -\frac{2}{\gamma(\phi)}\tilde{e}_{I}\tilde{e}_{J}R^{KL}(\omega) -\frac{\Lambda(\phi)}{6}\epsilon_{IJKL}\tilde{e}^{I}\tilde{e}^{J}\tilde{e}^{K}\tilde{e}^{L}\right) \nn\\
& + {\cal O}(\phi,R^2)  \label{einact1}
\end{align}
where ${\cal O}(\phi,R^2)$ denotes terms involving $\phi$ and terms \emph{quadratic} in the curvature tensor and where $\tilde{e}^{I}$ is conformally related to $e^{I}$ via a $\phi$-dependent factor such that the action (\ref{einact1}) is in the Einstein frame i.e. with no $\phi$-dependent coefficients of the Palatini term.
If the ${\cal O}(\phi,R^2)$ has negligible effect on the equations of motion then it is known \cite{Westman:2013mf} that equations of motion resulting from (\ref{einact1}) are equivalent to those of a second-order scalar-tensor theory of a metric $\tilde{g}_{\mu\nu}$ minimally coupled to a scalar field $\tilde{\phi}(\phi)$ possessing a canonical kinetic term and with potential $V(\tilde{\phi})$. Such a system corresponds precisely to that of single scalar-field inflation and whether the field $\tilde{\phi}$ is a suitable candidate for the growth of cosmic structure depends on the exact form of $V(\tilde{\phi})$. The following steps are then necessary to enable a similar confrontation to the data as that in inflationary models:

\begin{itemize}
	\item It must be shown that the terms ${\cal O}(\phi,R^2)$ in this regime are either negligible or - if not - lead to correspondence with specific, different models of inflation.
	\item In the former case, an informed estimate of the expected form of $V(\tilde{\phi})$ must be obtained. To do this, it will be necessary to deduce the form of coefficients $\{a_{i},b_{1},c_{1}\}$ in the action (\ref{act5}), specifically whether possible dependence on $W^{AB}$ can be ignored or not. Similar considerations likely come into play for the case where ${\cal O}(\phi,R^2)$ terms cannot be ignored. 
	\item Given the results of the previous item, it should then be possible to compute observables such as the tensor to scalar ratio, spectral indices of primordial perturbations and the extent to which primordial perturbations are adiabatic \cite{Mukhanov:1990me}. 
\end{itemize}
If one of these models seems promising as an inflaton candidate, there may be an advantage to the $SU(2,2)$-gravity approach to inflation in that the coupling of $\tilde{\phi}$ to matter is specifically prescribed by (\ref{wmatter}). This may make it simpler to come up with unambiguous predictions about the origin of structure in the universe via the decay of $\tilde{\phi}$ into matter and metric fields. Collectively, the results here will produce observable implications for anisotropies in the cosmic microwave background, potentially leading to strong constraints on the theory.

\subsection{Dark Matter}
In the standard model of cosmology, a sizeable contribution to the matter in the universe is due to \emph{dark matter} - a component that behaves as a near-pressureless, perfect fluid that has negligible coupling to known matter fields. A simple candidate for dark matter is a scalar field minimally coupled to gravity and with no coupling to matter; if the field is sufficiently massive then it will behave identically to dark matter for much of the universe's history \cite{Mukhanov:2005sc}. We have seen in Section \ref{perts} that $\phi$ can behave like a massive scalar field at the level of perturbations around simple backgrounds and we have seen from Section \ref{chirmass} that the description of spinors necessarily involves - in the General-Relativistic limit - new, massive fermions unknown to the standard model of particle physics. Both $\phi$ and the new fermions couple to fields other than the gravitational fields $A^{AB}$ e.g. $\phi$ couples to all matter fields whilst the new spinors couple to $\phi$ and standard model gauge fields. If $\phi$ or the new fermions may be dark matter candidates it will be necessary to calculate whether the couplings to other fields prevent the fields from being effectively `near-pressureless' and possessing negligible coupling to known matter on cosmological scales.

\subsection{Dark Energy}
The observed late-time acceleration of the universe may be explained in the context of General Relativity by a cosmological constant. However, a great deal of research has looked into the possibility the acceleration may be due to new fields in the universe or modifications to gravity - approaches collectively referred to as \emph{dark energy} \cite{Clifton:2011jh}. We note that it can be shown that due to the polynomial nature of (\ref{act5}), that the term $\Lambda(\phi)$ in (\ref{einact1}) \emph{necessarily} depends on $\phi$. Therefore, effects resembling a cosmological constant must be due to $\phi$ dynamically reaching approximate constancy i.e. the explanation for late-time acceleration of the universe in the context of $SU(2,2)$ gravity is necessarily a dark energy based one. As with the possibilities of $\phi$ playing a role in inflationary or dark matter phenomenology, an informed calculation of expected forms of $\{a_{i},b_{1},c_{1}\}$ and the effect of couplings between matter and gravity will be necessary to make meaningful contact with data. For example, it is conceivable that $\phi$ may vary with cosmic time as the universe evolves and this may lead to deviations an effective cosmological constant in the gravitational field equations; such deviations can be observable in data from supernovae and the cosmological microwave background \cite{DiValentino:2017zyq,Tutusaus:2017ibk}.

\section{Relation to other work}
\label{other}
We now discuss a number of alternative approaches that have been made to recovering gravitational theory from $SU(2,2)\simeq SO(2,4)$ gauge theories.

The work by Kaku et al. \cite{Kaku:1977pa} was discussed in some detail in Section \ref{weylrel};
a somewhat different approach was pursued by Kerrick \cite{Kerrick:1995tq} who introduced spacetime  scalar fields with $SU(2,2)$ index structure $Y_{\alpha\beta}= -Y_{\beta\alpha}$  and $Z_{\alpha\beta} = -Z_{\beta\alpha}$ (linear combinations of that author's original fields $I_{\alpha\beta}$ and $\iota_{\alpha\beta}$) alongside the $SU(2,2)$ connection $A^{\alpha}_{\ph{\alpha}\beta}$. Such fields were assumed to live in the six-dimensional space of matrices satisfying (\ref{selfdual}) and so are equivalent to introducing two $SO(2,4)$ vectors $\{Y^{A},Z^{A}\}$. The author implicitly constrained $Y^{A}Y_{A} = \mathrm{cst.} > 0$ and $Z^{A}Z_{A} = \mathrm{cst.} <0$, thus breaking $SU(2,2) \rightarrow SO(1,3)$. An action quartic in these fields and quadratic in the curvature $F^{AB}$  was shown to be equivalent to Einstein-Cartan gravity in the presence of a cosmological and Holst term.

Different again is the approach of Aros and Diaz \cite{Aros:2013yaa} who discuss an $SO(2,4)$ gauge theory on a \emph{five}-dimensional manifold. This is in some respects a higher-dimensional analogue of 
theories based on $SO(2,3)$ in four dimensions. The authors consider one of the five dimensions to have the topology of a circle and make the following ansatz for the \emph{five}-dimensional $A^{AB}\equiv A_{\mu}^{\ph{\mu}AB}dx^{\mu}+A_{y}^{\ph{y}AB}dy$:

\begin{eqnarray}
A^{AB} = \left(\begin{array}{ccc}0&\frac{1}{2}(e^I+f^I)&\xi d\phi\\-\frac{1}{2}(e^I+f^I)&\omega^{IJ}&-\frac{1}{2}(e^I-f^I)\\-\xi d\phi&\frac{1}{2}(e^I-f^I)&0\end{array}\right)
\end{eqnarray}
where fields are taken to be independent of coordinates along the circular dimension but may depend on the remaining four coordinates. The authors then propose a Chern-Simons-type action (an integral over a five-form) with explicit $SO(2,4)$ symmetry breaking (e.g. the $SO(1,4)|SO(2,3)$ invariant $\epsilon_{ABCDE}$ is used in the action instead of the $SO(2,4)$ invariant $\epsilon_{ABCDEF}$ without the use of a Higgs field to accomplish this covariantly). After dimensional reduction (integration of the action over the circle) it is found that the resulting four-dimensional action is that of Conformal Einstein-Cartan theory coupled to a scalar field $\xi(x)$. This theory can be related to fourth-order conformal gravity by adding constraints in the manner of Section \ref{weylrel} and freezing out the field $\xi(x)$. In higher dimensions yet, a novel approach to spacetime and gravity based on the group $SO(2,4)$ on eight-dimensional manifolds has been explored by Hazboun and Wheeler \cite{Hazboun:2013lra}.

\section{Discussion and conclusions}
\label{disc}
We now discuss the results contained in the paper and present our conclusions.

In this paper we have presented an approach to gravity as a spontaneously-broken gauge theory based on the group $SU(2,2)$ using the pair $\{A^{AB},W^{AB}\}$. We saw that when the degrees of freedom in the field $W^{AB}$ were `frozen' to a symmetry-breaking solution, that a locally scale-invariant generalization of the Einstein-Cartan gravity emerged; furthermore, when the two-frame fields $e^{I}$ and $f^{I}$ were aligned or anti-aligned then conformalized General Relativity was recovered i.e. General Relativity conformally coupled to a scalar field. It was then shown that fourth-order Weyl and conformal gravity could be kinematically embedded within the theory.

Next we focused on obtaining solutions for the theory when no constraints were placed on fields. Solutions corresponding to de Sitter space and anti-de Sitter space were found. Perturbations around these spacetimes were considered and the Lagrangian quadratic in these perturbations was constructed. It was found that the spectrum of perturbations was that of General Relativity, a massive scalar field, and a massless one-form field, each decoupled from one another at this level of perturbations. Thus from a gravitational theory possessing first-order field equations, three types of kinetic content yielding second-order field equations emerge in the perturbations:  perturbations identical to the metric perturbations of General Relativity for a field $\delta \theta^{I}$ with dynamics provided by $\delta \omega^{IJ}$;  scalar field perturbations described by a field $\delta\alpha$ with dynamics also provided by parts of $\delta \omega^{IJ}$; Weyl field perturbations $\delta c$ with dynamics provided by $\delta W^{IJ}$.
As noted at the end of Section \ref{perts}, the full theory may possess additional degrees of freedom and an important next step will be to show whether this is the case or not and how this affects the legitimacy of a perturbative approach to solutions where not all degrees of freedom have dynamics at linear order.

It is important to stress that these results, although encouraging, are provisional and that there is some way to go before it can be thoroughly demonstrated that a theory of gravity based on $SU(2,2)$ can reproduce the phenomenological success of General Relativity. In particular, we lack a clear understanding of whether there exists a convergence mechanism within the theory that somehow dynamically favours the General-Relativistic limit. Furthermore, from the perspective of the dynamical emergence of scale in the gravitational sector, do there indeed exist solutions which simultaneously contain regions where presumably no  scale can be defined which evolve to regions possessing notions of scale (e.g. field configurations approximate to the General-Relativistic limit)? For how long can no length exist?

Another central point of uncertainty seems to be that of the dynamics of $W^{AB}$. Though we found sets of parameters $\{a_{1},b_{1},c_{1}\}$ such that the `scalar field' part $\phi$ of $W^{AB}$ was static in our background solutions and had a positive effective mass-squared at the level of perturbations, it is not entirely clear this is necessary for agreement with experiment;  it is conceivable that the field may more naturally `roll' down an effective potential as the universe evolves and so possess time variation. Indeed, for gravity based on the groups $SO(1,4)|SO(2,3)$ it was found that Peebles-Ratra rolling quintessence was rather easily recovered \cite{Westman:2014yca,Peebles:1987ek}. 

We have additionally shown how it is possible to couple the $SU(2,2)$ gravitational fields to matter in polynomial Lagrangians that yield first-order field equations. As a consequence, time evolution of gravity and matter in the Lagrangian formulation is determined entirely by the values of fields at a moment in time. This brings to mind Zeno's `Arrow Paradox' wherein it is suggested that the trajectory of a moving arrow can be broken down into a series of moments, and in each moment the arrow is `static', perpetually contained within that moment. If the arrow is static in each moment, how does it move and how does it know where to move? In second-order Lagrangian theories, there is `hidden' information encapsulated in each moment: that of the velocity of the arrow. Moving to a Hamiltonian formulation, the arena of reality is that of phase space wherein the information present in the moment includes the arrow's position and momentum. From the perspective of first-order Lagrangian formulations of field theories, all this information is present in the field configuration space itself. For example, for the adjoint Higgs field we found in Section \ref{adjhiggs} that:

\begin{eqnarray}
\Phi^{AB} &\simeq &  \left( \begin{array}{cc}
({\cal F})^{IJ} & {\cal D}^{I}\Phi v^{a} \\
- {\cal D}^{I}\Phi v^{a} & \Phi \epsilon^{ab}
 \end{array}\right)
\end{eqnarray}
where ${\cal D}^{I}\Phi = (e^{-1})^{\mu I}{\cal D}_{\mu}\Phi$ etc. At any moment in time, the field $\Phi^{AB}$ contains the information about the value of the adjoint Higgs $\Phi$, how the Higgs field is changing (via the time derivative within the spacetime derivative ${\cal D}$), and also how the gauge field for the internal group that $\Phi$ is in the adjoint representation of is changing (via the field strength ${\cal F}$).

 In the table below we summarize the differences between gravity and how it couples to matter between Einstein-Cartan gravity (based on the group $SL(2,C)$) and the present model of gravity (based on the group $SU(2,2)$):

\begin{center}
    \begin{tabular}{| l | l | l | l | l | l | }
    \hline
    Model & Gravity & Fund. Higgs & Adjoint Higgs & Gauge Fields & Spinors \\ \hline
    $SL(2,C)$ & $\{e_{\mu}^{I},\omega_{\mu\ph{I}J}^{\ph{\mu}I}\}$ & $\varphi^{i}$ &$\phi^{i}_{\ph{i}j}$ & ${\cal B}_{\mu\ph{i}j}^{\ph{\mu}i}$ & $Weyl$ \\ \hline
    $SU(2,2)$ &$\{Y^{C\ph{D}A\ph{B}}_{\ph{C}D\ph{A}B},A_{\mu\ph{A}B}^{\ph{\mu}A}\}$& $\varphi^{Ai}$ &\multicolumn{1}{r}{$ \{\phi^{C\ph{D}i}_{\ph{C}D\ph{i}j},{\cal B}_{\mu\ph{i}j}^{\ph{\mu}i}\}$} && $Dirac$\\ 
    \hline
    \end{tabular}
\end{center}
where $i,j,k,\dots$ are indices in the fundamental representation of the matter sector symmetry group ${\cal G}$. From this perspective, the ingredients of gauge theories are a \emph{pair} $\{\phi^{A\ph{B}i}_{\ph{A}B\ph{i}j},{\cal B}_{\mu\ph{i}j}^{\ph{\mu}i}\}$ which from a second-order perspective describe Yang-Mills type dynamics of ${\cal B}$ alongside a massive adjoint Higgs field $\Phi$.  The field $\Phi$ in principle need not obtain a non-vanishing vev and so it need not break the gauge symmetry ${\cal G}$. It is interesting to note that gravity as a gauge theory couples to itself via precisely the same prescription: if we take the group ${\cal G}$ to be $SO(2,4)$ and identify $i,j,k,\dots$ with $A,B,C,\dots$ we recover the pair 
$\{\phi^{A\ph{B}C}_{\ph{A}B\ph{C}D},{\cal B}_{\mu\ph{A}B}^{\ph{\mu}A}\}$. Identifying $\phi^{A\ph{B}C}_{\ph{A}B\ph{C}D} = Y^{A\ph{B}C}_{\ph{A}B\ph{C}D}\equiv \frac{1}{2} \epsilon^{A\ph{B}C\ph{D}EF}_{\ph{A}B\ph{C}D}W_{EF}$ and ${\cal B}_{\mu\ph{A}B}^{\ph{\mu}A} = A_{\mu\ph{A}B}^{\ph{\mu}A}$ leads to the fields describing gravity.

Clearly a challenge for an $SU(2,2)$ description of gravity is the prediction of additional fermions beyond those in the standard model of particle physics: the fact that all spinor fields are Dirac spinors implies that there exist fermions with the same hypercharges but opposite handedness as the observed standard model fermions. 
We saw in Section \ref{chirmass} that generally the left and right handed parts of an $SU(2,2)$ spinor $\chi^{\alpha}$ will not have the same mass, but whether there exists a successful mechanism for making the theory compatible with experiment is an open issue\footnote{See \cite{BenTov:2015gra} for a discussion about recent proposals on what may be a similar issue in the context of the fate of non-observed additional generations of particles in $SO(18)$ grand unification schemes.}.

In conclusion, we have tried to argue in this paper that there is a firm experimental motivation for pursuing modified theories of gravity (the evidence for dark matter and cosmic inflation) and that a promising direction to take is to attempt to construct modifications of gravity based on larger symmetries than the local Lorentz invariance of General Relativity. We have argued that the new scalar and fermionic degrees of freedom in gravity and matter may have a cosmological role to play and we stress that a vital next step will be to construct unambiguous predictions from the model.

Finally, we now discuss some more speculative ideas based on the findings in the paper. Recall that the General-Relativistic limit of the Conformal Einstein-Cartan theory corresponded to the case where frames $e^{I}$ and $f^{I}$ were aligned or anti-aligned i.e. $f^{I} = \pm \Omega^{2}e^{I}$, and this determined the sign of the cosmological constant for the resulting theory (in the absence of matter). It is interesting to wonder whether there could exist solutions where $f^{I}$ varies smoothly (presumably passing through 0 along the way) from being aligned with $e^{I}$ to anti-aligned with $e^{I}$; what would be the interpretation of such solutions? This behaviour may also be possible in the full $\{A^{AB},W^{AB}\}$ theory.

In the unconstrained theory, though we have focused on finding maximally symmetric spacetime solutions (solutions with Lorentzian signature metric), in principle there may be very different phases of the theory depending on the symmetry breaking behaviour of the field $W^{AB}$. For instance, there can also exist forms of $W^{AB}$ that are preserved under $SO(4)\times SO(2)$, yielding a Euclidean theory of gravity with an additional local $SO(2)$ symmetry. It was found in the case of $SO(1,4)|SO(2,3)$ gravity a dynamic symmetry breaking field could transition between Lorentzian and Euclidean phases in simple cosmological models \cite{Magueijo:2013yya}. It would be interesting to see if this is also possible for the $SU(2,2)$ theory. If $W^{AB}=0$ then the entire $SO(2,4)$ symmetry may be preserved though it is not clear whether an interpretation of a `spacetime' theory is any longer possible here.

An important issue is whether big bang and and black hole singularities will still be present in this theory of gravity. It is quite possible that the new degrees of freedom beyond those in General Relativity (i.e. the Higgs field $W^{AB}$, and degrees of freedom in $A^{AB}$ outside of the General-Relativistic limit) would have a large influence in extreme situations and significantly modify the evolution. This would require the study of exact solutions to the full theory possibly with matter included.

Regarding the coupling of gravity to matter in this picture, we note that the matter actions that we considered were at most linear in the covariant derivatives $D\phi^{AB}$ and $D\varphi^{A}$. What would be the effect of considering polynomial actions quadratic or cubic in these derivatives? Why not include such terms? Would alternative second-order scalar field theories such as those contained in the Horndeski formalism \cite{Horndeski:1974wa,Copeland:2012qf,deRham:2014lqa,Mota:2010bs} be recoverable in some cases? Additionally, if there is a background where the field $\phi$ in $W^{AB}$ possesses a time dependence, the `metric'  $\eta_{AB}D_{\mu}W^{A}_{\ph{A}C}D_{\nu}W^{CB}$
picks up an additional dependence upon $\partial_{\mu}\phi \partial_{\nu}\phi$- the new metric being then disformally related to the one in which $\phi$ is static. Disformal couplings of scalar fields to matter have been investigated \cite{Koivisto:2008ak,Zumalacarregui:2010wj,Koivisto:2012za,Emond:2015efw,vandeBruck:2015tna} and it would be interesting to see if links can be made. Indeed, more generally it would be very useful to characterize the deviations from General Relativity that a theory based on the pair $\{A^{AB},W^{AB}\}$ may result in to move towards comparison with cosmological data. For example, from the perspective of cosmological phenomenology it would be interesting to see to what extent the theory could fit into the formalism described within \cite{Baker:2011jy,Baker:2012zs}.

Finally, given the attempt to describe gravity in the language of the theories of particle physics, it would also be interesting to see whether unification of gravity and the matter sector would be possible. For instance, the smallest groups containing $SU(2,2)\simeq SO(2,4)$ and a grand unification group $SO(10)$ as commuting subgroups are $SO(4,12)$ and $SO(2,14)$ \cite{Lisi:2010uw}.  Alternatively, the smallest group containing $SU(2,2)$ and $SU(5)$ as commuting subgroups is $SU(2,7)$.
Denoting indices in the adjoint representation of the `total unification' group as ${\cal A},{\cal B},{\cal C},\dots$ we can speculate that the gravity/gauge fields part of the theory would be described by a pair $\{C^{{\cal A}},\Phi^{{\cal A}{\cal B}}\}$: a connection $C^{{\cal A}} \equiv C_{\mu}^{\ph{\mu}{\cal A}}dx^{\mu}$ (containing $A_{\mu\ph{A}B}^{\ph{\mu}A}$ and ${\cal B}_{\mu\ph{i}j}^{\ph{\mu}i}$)  and a scalar Higgs field $\Phi^{{\cal A}{\cal B}}$ (containing $Y^{A\ph{B}C\ph{D}}_{\ph{A}B\ph{C}D}$ and $\phi^{A\ph{B}i}_{\ph{A}B\ph{i}j}$), from which Lagrangians polynomial in $\Phi^{{\cal A}{\cal B}}$, its covariant derivative, and the curvature of $C^{{\cal A}}$ can be constructed\footnote{In the case of $SO(4,12)$ there are a number of interesting symmetry breaking possibilities with the field $\Phi^{{\cal A}{\cal B}}=\Phi^{{\cal B}{\cal A}}$ including $SO(4,12)\rightarrow SO(1,3)\times SO(1,3)\times SO(1,3)\times SO(1,3)$, which may more resemble a theory containing four independent copies of the Einstein-Cartan field content.}. This is somewhat different than approaches based on $spin(3,11)\simeq SO(3,11)$ which contain $SL(2,C)\simeq SO(1,3)$ and $SO(10)$ as commuting subgroups \cite{Percacci:1984ai,Nesti:2009kk,Percacci:2009ij,Lisi:2010td}; these approaches typically involve fields aside from the $SO(3,11)$ connection possessing spacetime indices e.g. a co-tetrad in the fundamental representation of $SO(3,11)$ \cite{Percacci:1984ai} or a field which can dynamically tend to a Hodge dual operator on two-forms \cite{Lisi:2010td}.

\section*{Acknowledgements}
We thank James Bjorken, Johannes Noller, Jeffrey Hazboun, Andrew Randono, Andrew Tolley, Garrett Lisi, and Roberto Percacci for very useful discussions on first-order theories of gravity and unification. HW is indebted to Julian Barbour, Sean Gryb, Tim Koslowski, Flavio Mercati, Sebastiano Sonego, and Rafael Sorkin for many discussions about scale invariance in physics. We furthermore thank anonymous referees for comments and suggestions which have led to substantial improvement of the manuscript. Some of the research leading to these results has received funding from the European Research Council under the European Union's Seventh Framework Programme (FP7/2007-2013) / ERC Grant Agreement n. 617656 ``Theories and Models of the Dark Sector: Dark Matter, Dark Energy and Gravity.''

\bibliographystyle{hunsrt}
\bibliography{references}

\appendix

\section{Gauge approaches to gravity}
\label{gaugegravity}
In this appendix we provide a brief overview of attempts to formulate the classical theory of gravity as a gauge theory. 
As was discovered by Cartan, an elegant reformulation of General Relativity is provided by introducing a new, independent field $\omega^{I}_{\ph{I}J} \equiv \omega_{\mu\ph{I}J}^{\ph{\mu}I} dx^{\mu}$ - the \emph{spin connection} -  alongside $e^{I}$ in the description of gravity.  In these variables, the action for gravity is given by the Palatini action:

\begin{eqnarray}
S_{P}[\omega,e] &=& \int \epsilon_{IJKL} e^{I} e^{J} d\omega^{KL}+ \epsilon_{IJKL} e^{I} e^{J}\omega^{K}_{\ph{K}M}\omega^{ML} \label{pal}
\end{eqnarray}
where $\epsilon_{IJKL}$ is the completely antisymmetric Lorentz tensor invariant under $SO(1,3)$ transformations and multiplication of differential forms with one another is via the wedge product. Gravity from this perspective is known as Einstein-Cartan gravity\footnote{Within Einstein-Cartan gravity one may additionally consider polynomial terms  $\epsilon_{IJKL}e^{I}e^{J}e^{K}e^{L}$ and $e_{I}e_{J}(d\omega^{IJ}+\omega^{I}_{\ph{I}K}\omega^{KJ})$ which correspond to the cosmological constant term and the Holst term respectively. The Holst term only produces a non-zero contribution to field equations when the spin connection couples to other fields.}. Indeed, we can see that the terms containing $\omega^{I}_{\ph{I}J}$ in (\ref{pal}) combine to form the curvature two-form $R^{I}_{\ph{I}J}\equiv d\omega^{I}_{\ph{I}J}+\omega^{I}_{\ph{I}K}\omega^{K}_{\ph{K}J} $ of a non-Abelian gauge field and if $\omega^{IJ}$ is assumed to transform as a gauge field under $SO(1,3)$ transformations then the Lagrangian in (\ref{pal}) is manifestly $SO(1,3)$-invariant. 

The equation of motion for $\omega^{I}_{\ph{I}J}$ is as follows:

\begin{eqnarray}
de^{I}  +  \omega^{I}_{\ph{I}J} e^{J} &=& 0 \label{tor}
\end{eqnarray}

Unlike the equations of motion of gauge fields in particle physics, this equation is algebraic in $\omega^{I}_{\ph{I}J}$. One may solve (\ref{tor}) for $\omega^{I}_{\ph{I}J}(e)$; using this solution in the $e^{I}$ equation of motion yields the Einstein field equations. Alternatively, the solution for $\omega^{I}_{\ph{I}J}(e)$ may be inserted back into the action, resulting in the Einstein-Hilbert action of General Relativity, which - upon the addition of a boundary term - yields the Einstein field equations upon variation. Notably, however, using $\omega^{I}_{\ph{I}J}$ as an independent field allows one to write the Lagrangian for a fermionic field coupled to gravity polynomially in $\{e^{I},\omega^{IJ}\}$ and the fermion field. Due to the coupling to $\omega^{I}_{\ph{I}J}$, fermionic currents can act as a source term in the $\omega^{IJ}$ equation of motion; one may still solve for $\omega^{I}_{\ph{I}J}$ algebraically but it now depends on $e^{I}$ and the fermionic fields. Inserting the solution back into the action yields terms quartic in the fermionic fields, over and above terms usually present when describing gravity coupled to fermions in General Relativity \cite{Dolan:2009ni,spin,Diakonov:2011fs,Freidel:2005sn}.

Thus, the Einstein-Cartan approach results in a simplification of some actions involving the gravitational field and introduces structure (the $SO(1,3)$ gauge field $\omega^{I}_{\ph{I}J}$) reminiscent of the Yang-Mills fields of particle physics. However, the one-form $e^{I}$ has no counterpart amongst non-gravitational fields and its non-polynomial coupling to matter scalar and gauge fields introduces non-linearity into the field equations for $e^{I}$. Of course, such non-linearity may be an essential part of gravitation; a popular modification to the Einstein-Cartan gravity theory has been to allow terms non-polynomial in $e^{I}$ into the action of pure-gravity, much as they are present in the Einstein-Cartan matter sector. This allows one to construct non-topological terms quadratic in the $SO(1,3)$ curvature $R^{IJ}(\omega)$. This approach typically is referred to as \emph{Poincar\'{e} gauge theory} and it presents a wealth of new phenomenology in the gravitational sector, notably the propagation of $\omega^{IJ}$ itself via its own field equations (as opposed to simply being solvable for $\omega^{IJ} = \omega^{IJ}(e)$ as in the Einstein-Cartan case) \cite{Ho:2015ulu,Obukhov:2006gea,Hehl:2013qga,Chen:2015vya,Cianfrani:2015yya}. 

Intriguingly though, there exists a re-writing of the Palatini action with appears a step closer to commonality with the ingredients of the gauge theories of particle physics. The idea, originally due to MacDowell and Mansouri \cite{MacDowell:1977jt} (closely resembling earlier work due to Cartan \cite{Wise:2006sm}), is to enlarge the gauge group of gravity from $SO(1,3)$ to $SO(2,3)$ or $SO(1,4)$ (henceforth collectively referred to as $SO(1,4)|SO(2,3)$).
The trick is to imagine there there exists structure in a hypothetical $SO(1,4)|SO(2,3)$ gauge theory to break the symmetry down to the $SO(1,3)$ of the Einstein-Cartan theory. For example, this could be accomplished via a gravitational Higgs field $V^{A}$ in the fundamental representation of $SO(1,4)|SO(2,3)$ achieving a non-vanishing expectation value for its norm $V^{2}\equiv \eta_{AB}V^{A}V^{B}$ \footnote{Where in the case of $SO(1,4)$, $\eta_{AB}$ is the invariant matrix $\mathrm{diag}(-1,1,1,1,1)$ and for $SO(2,3)$, $\eta_{AB}$ is the invariant matrix $\mathrm{diag}(-1,-1,1,1,1)$.}. This norm should be positive/spacelike for $SO(1,4)$ and negative/timelike for $SO(2,3)$. Then we may choose a gauge where $V^{A} =\ell \delta^{A}_{4}$ (where $\ell$ is a constant); the residual gauge transformations that leave this explicit form of $V^{A}$ invariant are those of $SO(1,3)$, and we may decompose the $SO(1,4)|SO(2,3)$ gauge field $A^{A}_{\ph{A}B}   \equiv A_{\mu\ph{A}B}^{\ph{\mu}A} dx^{\mu}$ as follows:

\begin{equation}
A^{A}_{\ph{A}B} \overset{*}{=} \left( \begin{array}{cc}
A^{I}_{\ph{I}J} & A^{I}_{\ph{I}4} \\
A^{4}_{\ph{4}I} &0
\end{array} \right)
\end{equation}
where the $\overset{*}{=}$ represents an equality that holds in a specified gauge. We see then that there is a new field in the formalism: $A^{I}_{\ph{I}4}$. This is a one-form field that will transform homogeneously under the residual $SO(1,3)$ transformations, precisely like $e^{I}$! There exists a simple action principle due to Stelle and West \cite{Stelle:1979aj} that - post symmetry breaking - contains the Palatini action. For concreteness we look at the case of the group $SO(1,4)$ and the action is given by:

\begin{eqnarray}
S_{SW}[A,V,\lambda] &=&   \int \alpha\epsilon_{ABCDE}V^{E}F^{AB}F^{CD} + \lambda (V_{A}V^{A} - \ell^{2})
\end{eqnarray}
where, $F^{AB}\equiv dA^{AB}+A^{A}_{\ph{A}C}A^{CB}$,  $\alpha$ is a constant and the four-form field $\lambda$ is introduced entirely to enforce the constraint that $V_{A}V^{A}$ is constant. To illustrate the relation of this action to Einstein-Cartan theory, we may enforce the fixed-norm constraint at the level of the action, choosing a gauge where $V^{A} = \ell\delta^{A}_{4}$, and identifying $A^{I}_{\ph{I}J} = \omega^{I}_{\ph{I}J}$, $A^{I}_{\ph{I}4} = e^{I}/\ell$ we have:

\begin{equation}
A^{AB} \overset{*}{=} \left( \begin{array}{cc}
\omega^{IJ} & \frac{e^{I}}{\ell} \\
-\frac{e^{I}}{\ell} &0
\end{array} \right) \label{aab1}
\end{equation}

\begin{eqnarray}
S_{SW}[\omega, e] \overset{*}{=} -\frac{2\alpha}{\ell}\int \epsilon_{IJKL}\left(e^{I}e^{J}R^{KL}- \frac{1}{2\ell^{2}}e^{I}e^{J}e^{K}e^{L}-\frac{\ell^{2}}{2}R^{IJ}R^{KL}\right) \label{ssw}
\end{eqnarray}
where the notation $\overset{*}{=}$ means that something holds in a specified gauge (here the gauge where $V^{A} = \ell \delta^{A}_{4}$. We see then that the Palatini action (plus a cosmological constant term and a topological term quadratic in $R^{IJ}$) can be recovered from a spontaneously-broken gauge theory\footnote{See \cite{Randono:2010cq} for a more detailed discussion of these steps towards regarding gravity as a gauge theory.}. 

The four-form $\lambda$ is a simple way to achieve a non-vanishing norm for $V^{A}$ but is not necessary as an ingredient of the theory. Indeed, it has been found that there exist polynomial actions solely in terms of the set $\{ A^{A}_{\ph{A}B},V^{A}\}$ that dynamically yield a non-vanishing expectation value of $V^{2}$ \cite{Chamseddine:1977ih,Westman:2013mf}. These theories possess a rich phenomenology with the dynamics of the scalar  $V^{2}$ acting as a potential source of cosmic inflation or quintessence, even facilitating more exotic behaviour such as cosmological changes of the signature of the four-dimensional metric \cite{Westman:2014yca}. 

\begin{figure}
	\begin{center}
		
		%% Nodes 
		\begin{tikzpicture}[node distance=2cm]
		\node (aw) [blob2] {$SO(1,4)|SO(2,3)$ \\ gauge theory \\
			$S[A,V]$};
		\node(sol)[blob5, right of = aw, xshift = 5.3cm]{Scalar tensor theory \\
			with Euclidean \\
			and Lorentzian phases};
		\node (cec) [blob3, below of=aw, yshift = -0.8cm] {Einstein-Cartan \\ theory \\ $S[\omega,e]$};
		\node (gr) [blob3, right of=cec,xshift = 5cm] {General Relativity \\ $S[e]$};
		
		%% Arrows
		
		\draw [dotted,arrow] (aw) --node[anchor = south]{$V$ unconstrained}(sol);
		\draw [arrow] (aw) --node[anchor = east]{Constrain  $V$} (cec);
		\draw[arrow](cec) --node[anchor = north]{Solve for $\omega(e)$}(gr);
		
		\end{tikzpicture}
		
	\end{center}
	
	\captionsetup{width=0.89\textwidth, font = small}
	\caption{Diagram depicting known results of $SO(1,4)|SO(2,3)$ gravity and its relation to General Relativity. The dotted path is a path taken with $V^{A}$ entirely unconstrained.}
	\label{swg}
\end{figure}
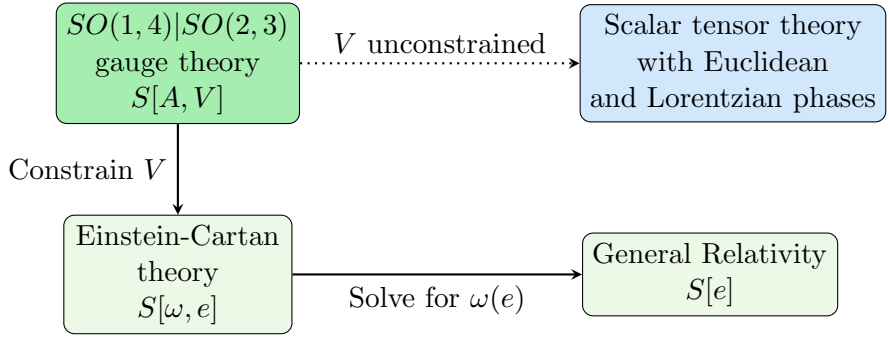

Therefore General Relativity (and scalar-tensor extensions thereof) can arise as a limit of a spontaneously-broken gauge theory based on $SO(1,4)|SO(2,3)$. However, the presence of a new degree of freedom $V^{2}$ in gravitation means the theory is more general than a re-casting of the Einstein-Cartan theory. 
This situation is summarized briefly in Figure \ref{swg}. Research into the link between $SO(1,4)|SO(2,3)$ groups and gravity and matter is ongoing \cite{Ikeda:2009xb,Kerr:2014sma,Kerr:2015tna,Jennen:2014mba,Palumbo:2015laa,Jennen:2015bxa}.

%% Flow diagram

There are, however, issues with a description of gravity based on $SO(1,4)|SO(2,3)$. Although it is possible to couple matter fields to gravity in a fashion 
consistent with the gauge principle \cite{Westman:2012zk}, we nevertheless encounter a somewhat unnatural structure when coupling gravity
to other Yang-Mills fields - specifically that auxiliary fields must be introduced to maintain the polynomial nature of the Lagrangian coupling gravity to Yang-Mills fields. In this paper we consider an alternative to the above $SO(1,4)|SO(2,3)$ approach wherein a solution to the problem of auxiliary fields is proposed.

\section{Useful identities}
\label{useful}

\begin{eqnarray}
\epsilon_{IJKL}e^{I}e^{J}e^{K}e_{N}\Phi^{LM}\Phi_{M}^{\ph{M}N} &=& -\frac{1}{8}\epsilon_{IJKL}e^{I}e^{J}e^{K}e^{L}\Phi^{MN}\Phi_{MN} \nn \\
  \epsilon_{IJKL}e^{I}e^{J}e^{K}e_{M}Y^{L}Y^{M} &=&  \frac{1}{4}\epsilon_{IJKL}e^{I}e^{J}e^{K}e^{L}Y^{M}Y_{M}\nn \\
 \epsilon_{IJKL}e^{I}e^{J}e^{K}Y^{L}{\cal D}\Phi &=&  \frac{1}{4}\epsilon_{IJKL}e^{I}e^{J}e^{K}e^{L}Y^{M}{\cal D}_{M}\Phi \nn \\
  \quad \epsilon_{IJKL}e^{I}e^{J}e^{K}\Phi^{LM}{\cal D}Z_{M} &=& \frac{1}{4} \epsilon_{IJKL}e^{I}e^{J}e^{K}e^{L}\Phi^{MN}{\cal D}_{M}Z_{N}\nn
  \end{eqnarray}

\section{Solving for $f$}
\label{solvf} 

We start from the action
\begin{align}
S[\omega,e,f,c]=\int \alpha \epsilon_{IJKL}e^If^JR^{KL}+\beta \epsilon_{IJKL}e^Ie^Jf^Kf^L+\gamma e^If_I dc
\end{align}
where $\{\alpha,\beta,\gamma\}$ are constants. Variation of the action yields the equations of motion
\begin{align}
\omega:\qquad 0&=\epsilon_{IJKL}d_\omega(e^Kf^L)\\
e:\qquad 0&=\alpha \epsilon_{IJKL}f^JR^{KL}+2\beta \epsilon_{IJKL}e^Jf^Kf^L+\gamma f_I dc\\
f:\qquad 0&=\alpha \epsilon_{IJKL}e^JR^{KL}+2\beta \epsilon_{IJKL}f^Je^Ke^L-\gamma e_I dc\\
c:\qquad 0&= d(e^If_I)
\end{align}
Solving the $f$-equation we obtain
\begin{align}
f^I=-\frac{3\alpha}{\beta}\left(R^I-\frac{R}{6}e^I\right)-\frac{\gamma}{\beta}e^I\lrcorner *dc.
\end{align}
Since $f^I$ appears algebraically in the action it is permissible to substitute this solution back into the action i.e. this operation does not alter the space of solutions. Carrying out this step yields
\begin{align}
S&=\frac{1}{\beta}\int -\alpha \epsilon_{IJKL}e^I\left(3\alpha R^J-\frac{\alpha R}{2}e^J+\gamma e^J\lrcorner *dc\right)R^{KL}-\gamma e_I\left(3\alpha R^I-\frac{\alpha R}{2}e^I+\gamma e^I\lrcorner *dc\right) dc\nn\\
&\qquad +\epsilon_{IJKL}e^Ie^J\left(3\alpha R^K-\frac{\alpha R}{2}e^K+\gamma e^K\lrcorner *dc\right)\left(3\alpha R^L-\frac{\alpha R}{2}e^L+\gamma e^L\lrcorner *dc\right)\nn\\
&=\frac{1}{\beta}\int -\alpha^2 \epsilon_{IJKL}e^I\left(3R^J-\frac{R}{2}e^J+\frac{\gamma}{\alpha} e^J\lrcorner *dc\right)R^{KL}- 3\alpha \gamma e_I R^Idc\nn\\
&\qquad+9\alpha^2\epsilon_{IJKL}e^Ie^J\left(R^K-\frac{R}{6}e^K\right)\left(R^L-\frac{R}{6}e^L\right)+6\alpha\gamma\epsilon_{IJKL}e^Ie^J\left(R^K-\frac{R}{6}e^K\right)e^L\lrcorner *dc\nn\\
&\qquad+\gamma^2\left(\epsilon_{IJKL}e^Ie^Je^K\lrcorner *dce^L\lrcorner *dc-2dc*dc\right)
\end{align}
where $*$ is the Hodge star operator on differential forms, built from $e^{I}$ and its matrix inverse, and $\lrcorner$ denotes the interior product between a vector and a differential form; in this notation $e^{I}\lrcorner$ means that the vector in question is $e^{\mu I}$. Using the fact that $e^I\lrcorner *H=\frac{1}{2}\epsilon^I_{\ph IJKL}e^J H^{KL}$ with $H\equiv dc$ let us now take a closer look at the penultimate term:
\begin{align}
&\epsilon_{IJKL}e^Ie^Je^K\lrcorner *dce^L\lrcorner *dc=\frac{1}{4}\epsilon_{IJKL}e^Ie^Je^Me^P \epsilon^K_{\ph KMNO} \epsilon^L_{\ph LPQR}H^{QR}H^{NO}\nn\\
&\sim\frac{e}{4}\epsilon_{IJKL}\varepsilon^{IJMP}\epsilon_{KMNO} \epsilon_{LPQR}H^{QR}H^{NO}=\frac{e}{2}(\delta_K^M\delta_L^P-\delta_K^P\delta_L^M)\epsilon^K_{\ph KMNO} \epsilon^L_{\ph LPQR}H^{QR}H^{NO}\nn\\
&=\frac{e}{2}\delta_K^M\delta_L^P\epsilon^K_{\ph KMNO} \epsilon^L_{\ph LPQR}H^{QR}H^{NO}-\frac{e}{2}\delta_K^P\delta_L^M\epsilon^K_{\ph KMNO} \epsilon^L_{\ph LPQR}H^{QR}H^{NO}\nn\\
&=\frac{e}{2}\epsilon_{KLNO} \epsilon^{KL}_{\ph{KL}QR}H^{QR}H^{NO}=2e\eta_{NR}\eta_{OQ}H^{QR}H^{NO}=-2eH^{IJ}H_{IJ}
\end{align}
Using $dc*dc\sim\frac{e}{2}H^{IJ}H_{IJ}$ we thus have 
\begin{align}
\epsilon_{IJKL}e^Ie^Je^K\lrcorner *dce^L\lrcorner *dc=-4dc*dc
\end{align}
and the action becomes
\begin{align}
S&=\frac{1}{\beta}\int -\alpha^2 \epsilon_{IJKL}e^I\left(3R^J-\frac{R}{2}e^J+\frac{\gamma}{\alpha} e^J\lrcorner *dc\right)R^{KL}- 3\alpha \gamma e_I R^Idc\nn\\
&\qquad+9\alpha^2\epsilon_{IJKL}e^Ie^J\left(R^K-\frac{R}{6}e^K\right)\left(R^L-\frac{R}{6}e^L\right)+6\alpha\gamma\epsilon_{IJKL}e^Ie^J\left(R^K-\frac{R}{6}e^K\right)e^L\lrcorner *dc\nn\\
&\qquad-6\gamma^2 dc*dc
\end{align}
Thus, the one-form $c$ looks like a gauge field with a standard Yang-Mills/Maxwell term proportional to $dc*dc$ alongside coupling of $dc$ to the curvature $R^{IJ}$.  We then have
\begin{align}
&\epsilon_{IJKL}e^I\frac{\gamma}{\alpha} e^J\lrcorner *dcR^{KL}=\frac{\gamma}{2\alpha}\epsilon_{IJKL}e^I\epsilon^J_{\ph JMNO}e^M H^{NO}  R^{KL}\nn\\
&=\frac{\gamma}{2\alpha}(\eta_{JO}\eta_{KM}\eta_{LN}+\eta_{JN}\eta_{KO}\eta_{LM}-\eta_{JO}\eta_{LM}\eta_{KN}-\eta_{JN}\eta_{LO}\eta_{KM})e^Je^M H^{NO}  R^{KL}\nn\\
&=\frac{\gamma}{\alpha}(\eta_{JO}\eta_{KM}\eta_{LN}+\eta_{JN}\eta_{KO}\eta_{LM})e^Je^M H^{NO}  R^{KL}\nn\\
&=\frac{2\gamma}{\alpha}H_{KJ}e^J e_L  R^{KL}= \frac{2\gamma}{\alpha}e_K\lrcorner dc DT^K
\end{align}
and also
\begin{align}
&6\alpha\gamma\epsilon_{IJKL}e^Ie^J\left(R^K-\frac{R}{6}e^K\right)e^L\lrcorner *dc=-3\alpha\gamma e^Ie^Je^M\epsilon_{LIJK}\epsilon^L_{\ph LMNO} H^{NO}\left(R^K-\frac{R}{6}e^K\right)\nn\\
&=6\alpha\gamma e^Ie^JH_{IJ}e_KR^K=12\alpha\gamma e_KR^K dc
\end{align}
where $D$ is the $SO(1,3)$ covariant derivative. The action now becomes 
\begin{align}
S&=\frac{1}{\beta}\int-3\alpha^2 \epsilon_{IJKL}e^I\left(R^J-\frac{R}{6}e^J\right)R^{KL}+9\alpha^2\epsilon_{IJKL}e^Ie^J\left(R^K-\frac{R}{6}e^K\right)\left(R^L-\frac{R}{6}e^L\right)\nn\\
&\qquad-2\alpha\gamma e_I\lrcorner dc R^{IJ}e_J +9\alpha\gamma e_KR^K dc-6\gamma^2 dc*dc \nn\\
 &= \frac{1}{\beta}\int \frac{\alpha^{2}}{4}\epsilon_{IJKL}{\cal C}^{IJ}{\cal C}^{KL} -6\gamma^2 dc*dc \nn \\
 &\qquad -2\alpha\gamma e_I\lrcorner dc R^{IJ}e_J +9\alpha\gamma e_KR^K dc-\frac{\alpha^{2}}{4}\epsilon_{IJKL}R^{IJ}
 R^{KL } \label{act9}
\end{align}
where the Weyl two-form has been defined as:

\begin{eqnarray}
{\cal C}^{IJ} \equiv  R^{IJ} -  6e^{I}\left(R^{J}-\frac{R}{6}e^{J}\right)
\end{eqnarray}
We observe that the relative sign of `Weyl-squared' term and Maxwell-type term in (\ref{act9}) are fixed. We can further develop (\ref{act9}) by noting that:

\begin{eqnarray*}
e_{I}\lrcorner (dc R^{IJ}e_{J}) &=& e_{I}\lrcorner (0) =0\\
&=& (e_{I}\lrcorner dc)R^{IJ}e_{J} + dc (e_{I}\lrcorner R^{IJ})e_{J} + dc R^{IJ}e_{I}\lrcorner e_{J}\\
&=&  (e_{I}\lrcorner dc)R^{IJ}e_{J} + dc R^{J}e_{J} 
\end{eqnarray*}
where we have used the fact that $R^{J}\equiv (e_{I}\lrcorner R^{IJ})$, $e_{I}\lrcorner e_{J}=\eta_{IJ}$, and $R^{IJ}\eta_{IJ}=0$. The action (\ref{act9}) can then be seen to reduce to:

\begin{align}
S&= \frac{1}{\beta}\int \frac{\alpha^{2}}{4}\epsilon_{IJKL}{\cal C}^{IJ}{\cal C}^{KL} -6\gamma^2 dc*dc +11\alpha\gamma e_KR^K dc-\frac{\alpha^{2}}{4}\epsilon_{IJKL}R^{IJ}R^{KL }
\end{align}

\end{document}